\documentclass[journal=jacsat,manuscript=article]{achemso}

\usepackage{graphicx}
\usepackage{tikz}
\usepackage{hyperref}
\usepackage{xr}
\usepackage{booktabs,caption}
\usepackage[flushleft]{threeparttable}

\usepackage[version=3]{mhchem} 
\usepackage{bm}
\usepackage{booktabs}

\author{Joseph D. Clark}
\affiliation[University of Illinois at Urbana-Champaign]
{School of Molecular and Cellular Biology,University of Illinois at Urbana-Champaign,Urbana, IL 61801, USA}
\author{Xuenan Mi}
\affiliation[University of Illinois at Urbana-Champaign]
{Center for Biophysics and Quantitative Biology,University of Illinois at Urbana-Champaign,Urbana, IL 61801, USA}
\author{Douglas A. Mitchell}
\affiliation{Department of Chemistry,University of Illinois at Urbana-Champaign,Urbana, IL 61801, USA}
\author{Diwakar Shukla}
\affiliation[University of Illinois at Urbana-Champaign]
{Center for Biophysics and Quantitative Biology,University of Illinois at Urbana-Champaign,Urbana, IL 61801, USA}
\alsoaffiliation{Department of Chemical and Biomolecular Engineering,University of Illinois at Urbana-Champaign,Urbana, IL 61801, USA}
\alsoaffiliation{Department of Bioengineering,University of Illinois at Urbana-Champaign,Urbana, IL 61801, USA}
\email{diwakar@illinois.edu}
\title[An \textsf{achemso} demo]
  {Substrate Prediction for RiPP Biosynthetic Enzymes via Masked Language Modeling and Transfer Learning}

\abbreviations{IR,NMR,UV}

\begin{document}

%
%
%
%
%

\begin{abstract}
Ribosomally synthesized and post-translationally modified peptide (RiPP) biosynthetic enzymes often exhibit promiscuous substrate preferences that cannot be reduced to simple rules. Large language models are promising tools for predicting such peptide fitness landscapes. However, state-of-the-art protein language models are trained on relatively few peptide sequences. A previous study comprehensively profiled the peptide substrate preferences of LazBF (a two-component serine dehydratase) and LazDEF (a three-component azole synthetase) from the lactazole biosynthetic pathway. We demonstrated that masked language modeling of LazBF substrate preferences produced language model embeddings that improved downstream classification models of both LazBF and LazDEF substrates. Similarly, masked language modelling of LazDEF substrate preferences produced embeddings that improved the performance of classification models of both LazBF and LazDEF substrates. Our results suggest that the models learned functional forms that are transferable between distinct enzymatic transformations that act within the same biosynthetic pathway. Our transfer learning method improved performance and data efficiency in data-scarce scenarios. We then fine-tuned models on each data set and showed that the fine-tuned models provided interpretable insight that we anticipate will facilitate the design of substrate libraries that are compatible with desired RiPP biosynthetic pathways.  
\end{abstract}

\section{Introduction}
\begin{figure}
    \centering
    \includegraphics[scale=1.28]{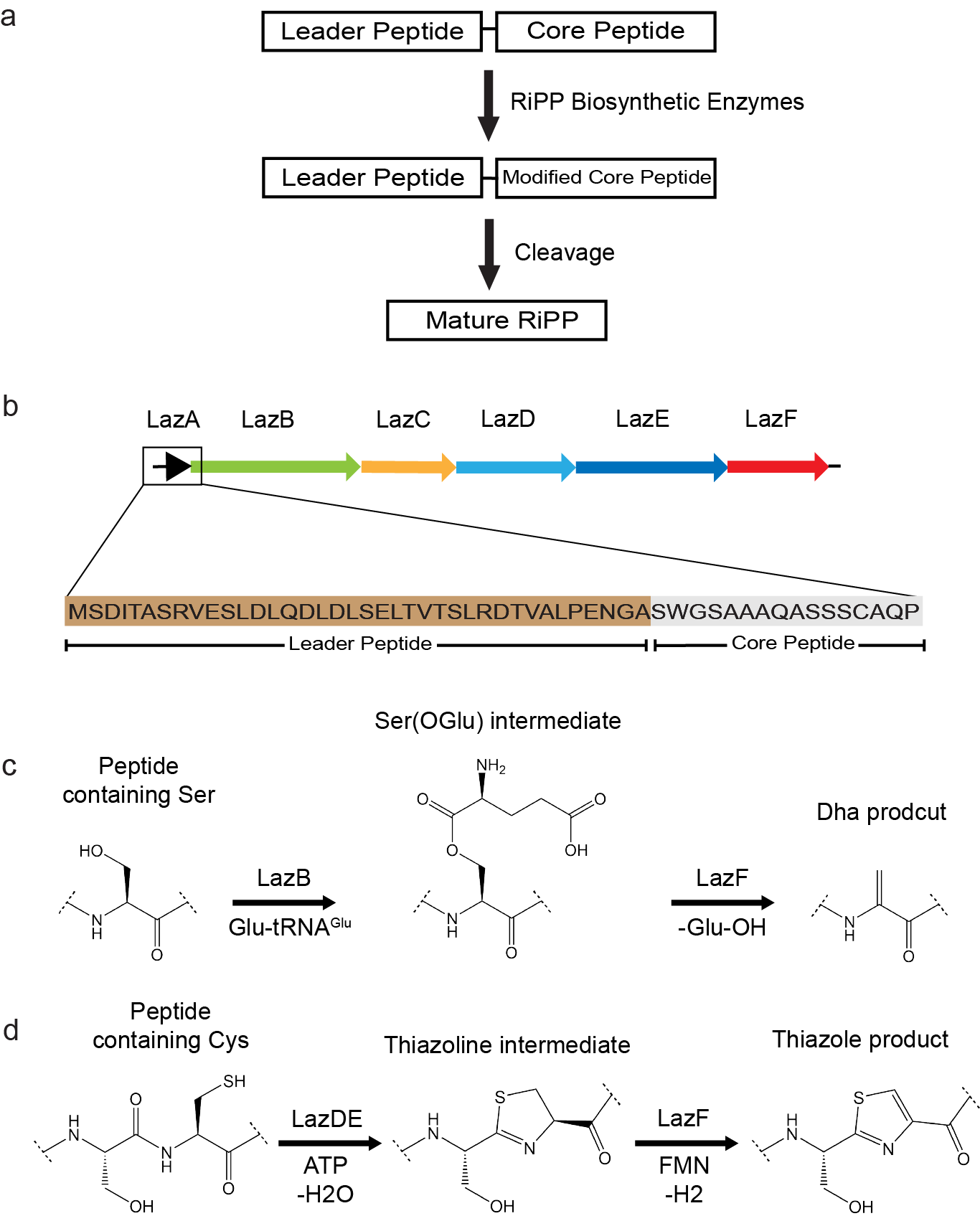}
    \caption{a) The generic biosynthesis pathway of RiPPs. RiPP precursor peptides contain a leader peptide and a core peptide. After  post-translational modifications in the core peptide, the leader peptide is cleaved. b) The lactazole biosynthetic gene cluster contains six proteins. LazA is the precursor peptide. LazB (tRNA-dependent glutamylation enzyme) and the eliminase domain of LazF form a serine dehydratase while LazD (RRE-containing E1-like protein)\cite{Burkhart2015}, LazE (YcaO  cyclodehydratase)\cite{Burkhart2017}, and the dehydrogenase domain of LazF comprise a thiazole synthetase. LazC is a pyridine synthase. c) Serine dehydration catalyzed by LazBF. d) Thiazole  formation catalyzed by LazDEF.}
    \label{fig:figure1}
\end{figure}

Ribosomally synthesized and post-translationally modified peptides (RiPPs) are a broad category of natural products with largely untapped clinical potential\cite{Ongpipattanakul2022, Fu2021}. A typical RiPP precursor peptide contains an N-terminal leader region followed by a core region (Figure \ref{fig:figure1})\cite{MontalbnLpez2021}. RiPP precursor peptides undergo post-translational modifications (PTMs) in the core region, which serve to restrict conformational flexibility, enhance proteolytic resistance, and chemically diversify the natural product\cite{MontalbnLpez2021}. After modification of the core peptide, the leader region is cleaved, releasing the mature RiPP. The PTMs are installed by RiPP biosynthetic enzymes, some of which display high levels of specificity while others act on diverse peptides\cite{Arnison2013}. A significant effort has been dedicated to characterizing the substrate preferences of RiPP biosynthetic enzymes and PTM enzymes in general, which, in many cases, cannot be explained by a simple set of rules\cite{Vinogradov2022, Ivry2017, Tang2014, Le2021, Song2021, Mahajan2021}. Consequently, machine learning and deep learning are increasingly used to develop predictive models of PTM specificity\cite{Meng2022, Yan2023, Vinogradov2022}. For instance, XGBoost was used to predict the protein substrates of phosphorylation and acetylation in multiple organisms,\cite{Smith2022} and a transformer-based protein language model was applied to predict glycation sites in humans\cite{Liu2022}. Finally, MusiteDeep is a web server for deep learning-based PTM site prediction and visualization for proteins\cite{Wang2020}.

Characterizing RiPP biosynthetic enzyme specificity is challenging, mainly due to the complexity of substrate fitness landscapes and the scarcity of sequences labeled as substrates or non-substrates\cite{Zhao2009, Kaltashov2012}. Accordingly, pretrained protein language models can be used to embed peptides as information rich vector representations to combat data scarcity\cite{Brandes2022}. Protein language models are transformer-based neural networks that learn the biological properties of polypeptides by predicting the identities of hidden residues in a training paradigm called masked language modeling\cite{Lin2023, https://doi.org/10.48550/arxiv.1906.08230}. Masked language modeling is a form of self-supervised learning, in which a model predicts features contained within the training data (e.g., masked residues) instead of experimentally determined property labels. The protein language model representations of polypeptide sequences, also called embeddings, can be extracted and used as feature vectors for training downstream machine learning models\cite{Rives2021, https://doi.org/10.48550/arxiv.2007.06225}. This is a canonical example of transfer learning, in which knowledge learned during one task is utilized in a distinct but related task\cite{https://doi.org/10.48550/arxiv.1911.02685, Shamsi2020}. Protein language model representations have seen widespread use in peptide prediction tasks such as antimicrobial activity and toxicity prediction\cite{Wang2023a, Wang2023b, Zhang2021, https://doi.org/10.48550/arxiv.2309.14404, Du2024}. However, protein language models have been trained mostly on protein sequences, which have are much larger and more structurally defined compared to peptides\cite{Fosgerau2015, Muttenthaler2021}. Therefore, protein language models may not fully capture peptide-specific features. Sadeh \textit{et al.}  trained self-supervised language models on peptide data, but unfortunately their models are not publicly available \cite{https://doi.org/10.48550/arxiv.2211.06428}. To the best of our knowledge, no self-supervised, sequence-based peptide language models are publicly available. Peptide prediction models may benefit from transfer learning paradigms in which protein language models are further trained on peptide data that is closely relevant to the downstream task. In a few cases, there exist large, high quality data sets characterizing the substrate specificity of specific RiPP biosynthetic enzymes\cite{Vinogradov2022, Huang2017}. In this work, we evaluated whether learning  such data sets in a self-supervised fashion could more effectively capture functional forms that are transferable to prediction tasks of other enzymes from the same biosynthetic pathway.

Transfer learning between the substrate preferences of enzymes from the same biosynthetic pathway could potentially enhance data efficiency and model performance in situations with low data availability. To date, little work has been performed to investigate transfer learning between substrate prediction tasks of related enzymes. Lu \textit{et al.} used a geometric machine learning approach to model the substrate preferences of protease enzymes\cite{Lu2023}. This work found that models trained to predict the substrates of a single protease were able to generalize to other protease variants with multiple amino acid substitutions. In the case of RiPP biosynthetic enzymes, transfer learning could also help evaluate the degree of shared features between distinct enzymes. Such insights could aid peptide engineering tasks and facilitate a more holistic understanding of RiPP biosynthesis.

Thiopeptides are a specialized form of pyritide antibiotics deriving mostly from Bacillota and Actinomycetota\cite{MontalbnLpez2021, Chan2020, Schwalen2018}. Lactazole A (LazA)\cite{Hayashi2014} is a natural product from the pyritide family of RiPPs\cite{Vinogradov2020a, Hudson2020} which is encoded by a biosynthetic gene cluster containing 5 synthetases (Figure \ref{fig:figure1}). A diverse array of precursor peptides can be converted to lactazole-like products by these biosynthetic enzymes which catalyze post-translational modifications\cite{Vinogradov2020b}. LazBF is a split Ser dehydratase which installs a Dha residue in LazA precursor peptides\cite{Vinogradov2021, Vinogradov2020d}. LazDEF is a split azole-forming enzyme complex which produces thiazoles in LazA precursor peptides\cite{Vinogradov2020c}. A previous study comprehensively profiled the peptide fitness landscapes of LazBF and LazDEF (LazC was not included in their study) via the generation of two data sets each containing over 8 million LazA core sequences labeled as substrates or non-substrates\cite{Vinogradov2022}. This study trained convolutional neural networks which showed excellent performance on substrate classification tasks. In the case of LazBF, dehydration sites and important residues were identified using integrated gradients\cite{pmlr-v70-sundararajan17a}, an interpretable machine learning technique which determines the positive or negative contribution of each input feature to the model's prediction. Despite the robust interpretability of their models, this study was unable to produce a general set of rules describing the substrate preferences of either LazBF or LazDEF. The comprehensive nature of the LazBF/DEF substrate data sets, and the fact that both data sets characterize related but distinct enzymes from the same biosynthetic pathway make them good candidates for exploring the plausibility of transfer learning between peptide substrate prediction tasks. 

\begin{figure}
    \centering
    \includegraphics[width=\textwidth]{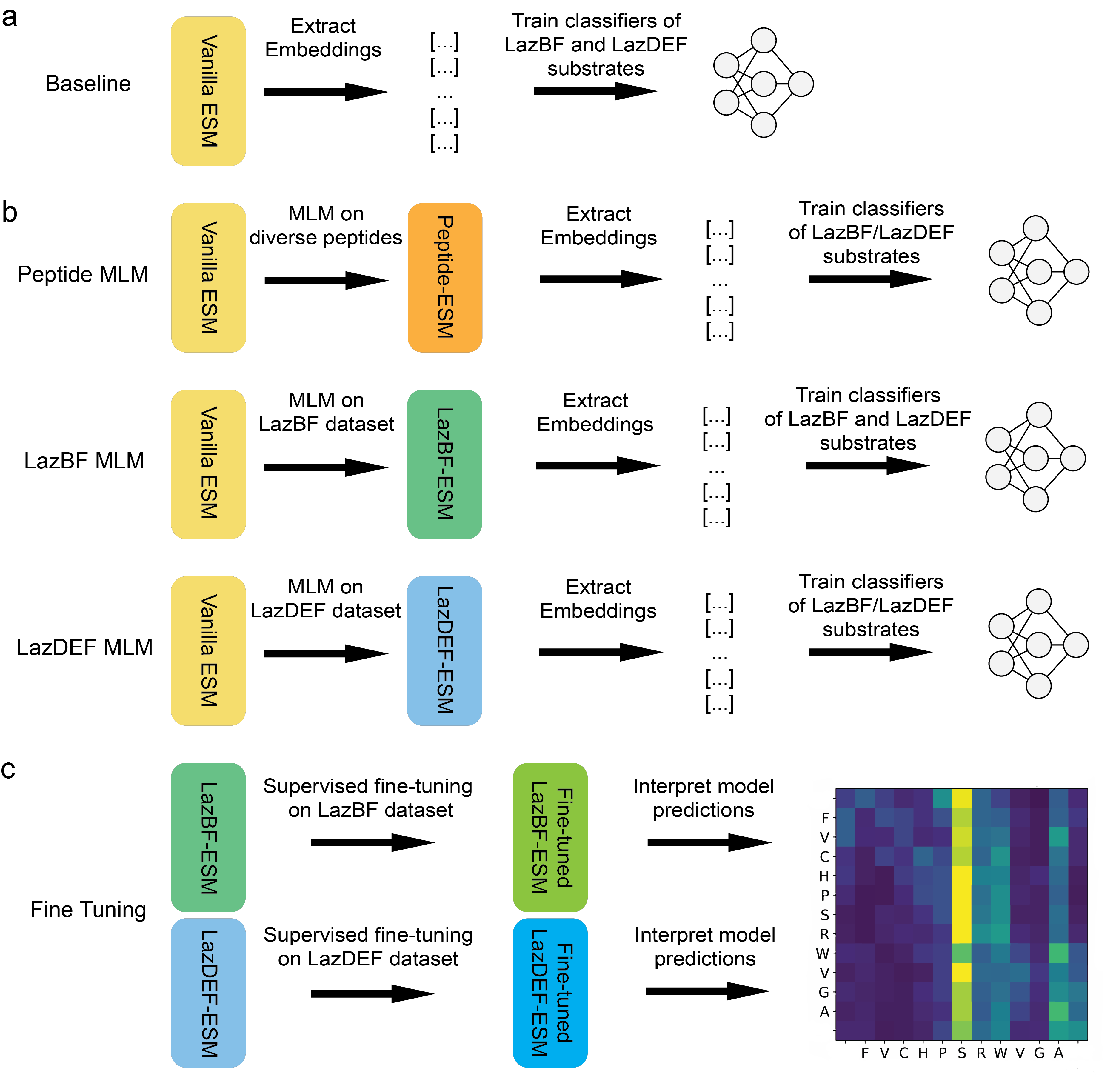}
    \caption{A schematic representation of the workflow for masked language modeling of LazBF and LazDEF substrate preferences. a) LazBF and LazDEF substrate/non-substrate embeddings were extracted from the protein language model ESM-2 (Vanilla-ESM). The baseline performance of downstream classification models was assessed. b) 3 copies of Vanilla-ESM were independently trained through masked language modeling of 3 peptide data sets. Embeddings were extracted and the performance of downstream classification models was compared to baseline. c) Models were further trained to directly classify LazBF/DEF substrates. The models’ predictions were analyzed with interpretable machine learning techniques including attention analysis (see methods).}
    \label{fig:figure2}
\end{figure}

In this work, we used masked language modeling to further train protein language models on RiPP biosynthetic enzyme substrates and non-substrates. We then evaluated transfer learning between the substrate preferences of LazBF and LazDEF. Specifically, we observed that embeddings from a self-supervised language model trained on LazBF substrates and non-substrates outperformed baseline protein language model embeddings on either substrate classification task. We show a similar result in the opposite direction, where embeddings from a self-supervised model of LazDEF substrates and non-substrates outperformed baseline embeddings on either substrate classification task. Embeddings from LazBF/DEF-specific language models also outperformed embeddings from a baseline peptide language model trained on a subset of PeptideAtlas\cite{Desiere2006}, a diverse data set of mass-spectrometry identified peptides. We then trained our language models to directly classify peptides as substrates or non-substrates through a process called fine-tuning. Finally, we evaluated the transfer of interpretable machine learning techniques between the LazBF and LazDEF substrate prediction tasks. Specifically, we showed that a model fine-tuned to classify LazDEF substrates correctly identified the residue types and positions important for LazBF substrate fitness. Figure \ref{fig:figure2} presents a schematic representation of our overall workflow. Our results suggest that 1) some degree of features are shared between the fitness landscapes of LazBF and LazDEF, and 2) masked language modeling and transfer learning lead to improved predictive performance on RiPP biosynthetic enzyme prediction tasks, especially when large unlabeled data sets are available. With the increasing power of high-throughput methods, this work could enable improvement on other substrate prediction tasks by leveraging large data sets and transfer learning. 

\section{Methods}


\subsection{Data Preprocessing}
\begin{figure}[t]
    \centering
    \includegraphics[width=\textwidth]{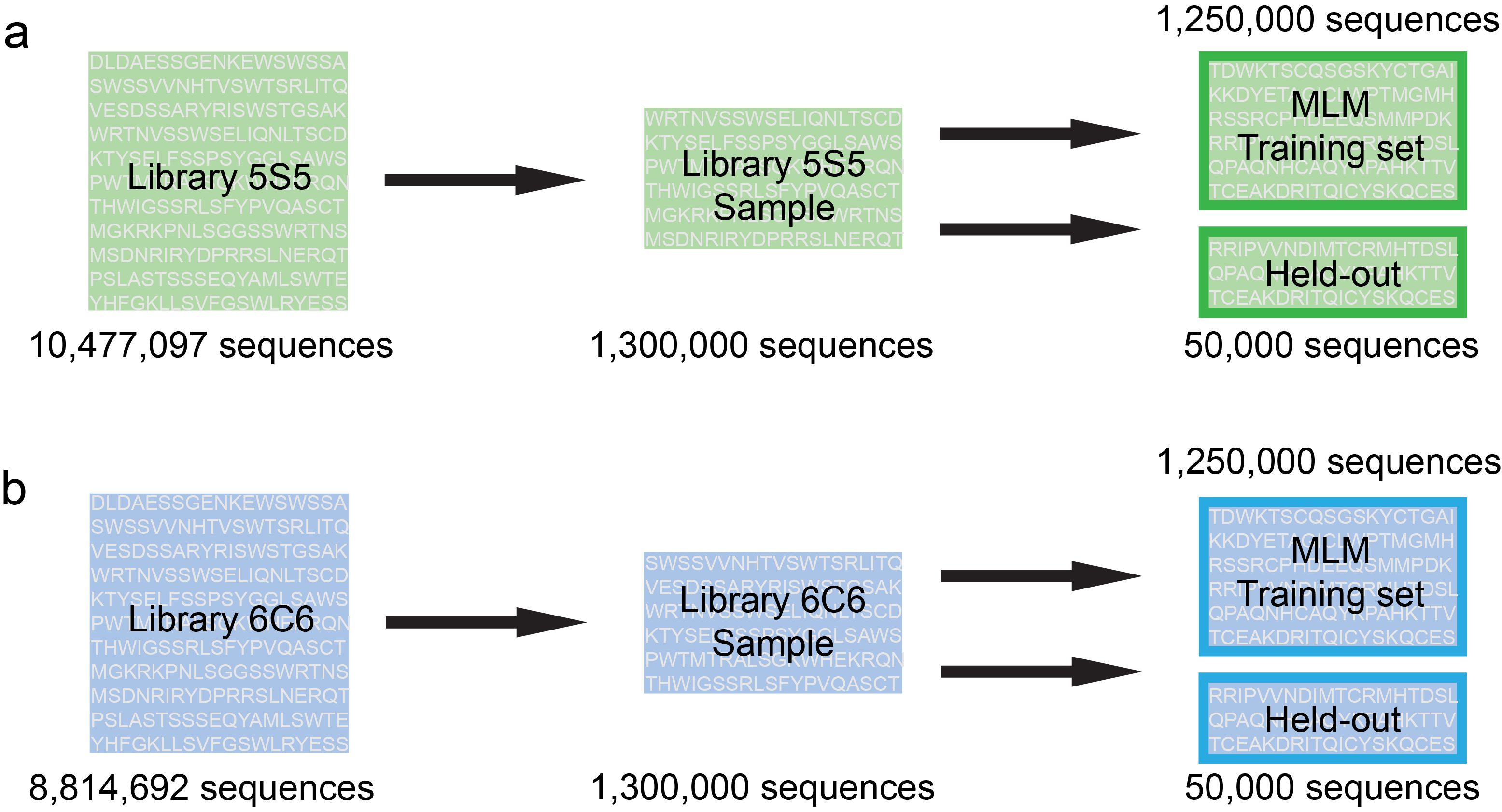}
    \caption{A schematic representation of our data preprocessing pipeline. a) LazA core sequences (\textit{n} = 1.3 million) were selected from library 5S5. A ‘held-out’ data set of 50,000 peptides was set aside for downstream model training and evaluation. b) LazA core sequences (\textit{n} = 1.3 million) were selected from library 6C6. A held-out data set of 50,000 peptides was set aside for downstream model training and evaluation.}
    \label{fig:figure3}
\end{figure}

Vinogradov \textit{et al.} used an mRNA display based profiling method and next-generation sequencing to generate two data sets of LazA core peptide sequences labeled as either substrates or non-substrates for LazBF and LazDEF respectively\cite{Vinogradov2022}. For LazBF substrates/non-substrates, each core peptide contained a serine residue flanked by five N-terminal and five C-terminal residues (library 5S5). For LazDEF substrates/non-substrates, each core region contained cysteine flanked by six residues on each side (library 6C6). Duplicate sequences were removed from both libraries. Pairs of identical sequences found in the substrate and non-substrate bins were removed. For both libraries, a sample of 1.3 million sequences containing an equal number of substrates and non-substrates was selected. A subset of 50,000 peptides from each sample was excluded as “held-out” data for training and validation of downstream models after masked language modeling. The remaining 1.25 million LazA core peptide sequences in each sample were used as the training data for masked language modeling. Importantly, none of the held-out sequences were seen during masked language modeling. Figure \ref{fig:figure3} provides a schematic of the data preprocessing pipeline.

In a later study, Chang \textit{et al.} used mRNA display based profiling to generate a data set of LazA core peptide sequences labeled as either substrates or non-substrates for the entire lactazole biosynthetic pathway (LazBCDEF)\cite{Chang2023}. This study comprehensively profiled the combined substrate preferences of all 5 synthetases as opposed to individual enzymes. This data set was preprocessed in a manner identical to the LazBF/DEF substrate data sets.

\subsection{Masked Language Modeling}
Masked language modeling is a widely-used strategy for pretraining large language models\cite{https://doi.org/10.48550/arxiv.1810.04805, Wang2023}. In the context of protein language models, masked language modeling takes a polypeptide sequence and replaces a random subset (15\%) of the amino acids with a masking token ([MASK]). Partially masked polypeptides are fed into the model, which is optimized to predict the identity of masked residues given the context of the surrounding amino acids. This `self-supervised' pretraining objective has enabled models to learn the biological features of proteins including secondary structure, long range residue-residue contacts, and mutational effects\cite{Rives2021}. We hypothesized that, for a pretrained protein language model, further masked language modeling of the LazBF or LazDEF substrate preference data sets would update the model's representations and enable better discrimination between substrates and non-substrates. Additionally, we sought to test how well the representations from a model trained on LazBF substrates and non-substrates would be able to discriminate LazDEF substrates and vice versa.

ESM-2 is a family of transformer-based protein language models with state-of-the-art performance on various protein and peptide prediction tasks\cite{Rives2021, Meier2021}. ESM-2 is composed of a series of encoder layers, where each layer takes a numerically represented polypeptide as input and maps it to a continuous vector representation. Layers are stacked sequentially to produce increasingly rich representations. A 12-layer, 35 million parameter version of ESM-2 was used as a baseline model (Vanilla-ESM). 3 copies of Vanilla-ESM underwent additional training using masked language modeling. ``LazBF-ESM" was trained on 1.25 million LazA core peptide sequences from the LazBF data set. ``LazDEF-ESM" was trained on 1.25 million LazA core peptide sequences from the LazDEF data set. ``Peptide-ESM" was trained on a random sample of 1.25 million sequences from Peptide Atlas\cite{Desiere2006}. Each model was trained for 1 epoch (i.e., one complete pass through the training data set) on their respective data sets with a learning rate of $3 \times 10^{-6}$ and a batch size of 512. 

\subsection{Embedding Extraction and Downstream Model Training}
Each layer of a protein language model produces vector representations of protein sequences that encode biological structure and function\cite{Rao2020, Bepler2021}. Protein language model representations, are commonly used as the input to downstream machine learning models trained on various protein and peptide prediction tasks\cite{Marquet2021, Weissenow2022}. The embeddings for all core peptides in the LazBF and LazDEF held-out data sets were extracted from Vanilla-ESM, Peptide-ESM, LazBF-ESM, and LazDEF-ESM. For each sequence, the last layer representation was obtained as a matrix of shape $L \times 480$, where $L$ was the length of the sequence. The last layer representation was averaged across the length dimension to obtain a single 480-dimensional mean representation. The embeddings from the held-out LazBF and LazDEF data sets were used for training and validation of various machine learning models as described in the proceeding subsections. Each downstream model type was trained and validated independently on both the LazBF and LazDEF held-out data sets. All downstream models were implemented in Scikit-learn\cite{JMLR:v12:pedregosa11a}. \texttt{StandardScaler} was applied to all embeddings following standard protocols prior to training.

\subsubsection{Unsupervised clustering}
Unsupervised \textit{k}-Means clustering was used to assess how well the distinction between substrates and non-substrates was represented in the high-dimensional ESM-2 embeddings. A \textit{k}-Means clustering model with \texttt{n\_clusters = 2} was fit to each set of embeddings. The accuracy between the ground truth labels and the \textit{k}-Means predicted labels was then calculated, along with the precision, recall, area under the receiver operating characteristic (AUROC), and F1 score. For each set of embeddings, five separate \textit{k}-Means models were trained with different \texttt{random\_state} parameters. The model's final performance was described by the average of the \textit{k}-Means metrics.

\subsubsection{Supervised classification models} 
Supervised learning models are trained by predicting properties of labeled data points (e.g., substrate or non-substrate). Logistic regression (LR), \textit{k}-nearest neighbors classifier (KNN), random forest (RF), AdaBoost (AB), support vector classifier (SVC), and multi-layer perceptron (MLP) models were trained via supervised learning to predict LazBF and LazDEF substrates using the embeddings from each of the 4 protein language models as input. All embeddings were reduced to 50 dimensions with principal component analysis (PCA) before being used as the input for supervised classification models. Stratified 5-fold cross validation was performed for each model. For each fold, the accuracy, precision, recall, AUROC, and F1 score between the ground truth labels and the predicted labels was calculated. The final model performance was described by the average metrics for all 5 folds. To emulate real-world scenarios in which training data is limited, each model type was trained and validated under 3 conditions. In the ``high-N" condition, 5-fold cross validation was performed such that for each fold, 10,000 peptides were used for validation, and 40,000 peptides were used for model training. In the ``medium-N" condition, 5-fold cross validation was performed such that for each fold, 10,000 peptides were used for validation, but only a random sample of 1,000 peptides were used for model training. In the ``low-N" condition, 5-fold cross validation was performed such that for each fold, 10,000 peptides were used for validation, but only a random sample of 100 peptides were used for model training. Hyperparameters of each supervised model were optimized separately for each set of embeddings under each condition using grid search. The optimized hyperparameters for all downstream models are in Tables S1-S3.

\subsubsection{Embedding space visualization} 
t-Distributed Stochastic Neighbor Embedding (t-SNE) was used to visualize the embeddings from each protein language model. A sample of 5,000 peptides from both held-out data sets were selected for visualization. The 480-dimensional embeddings were first reduced to 100 dimensions with PCA, and then further reduced to two dimensions with t-SNE.

\subsection{Fine-Tuning, Integrated Gradients, and Attention Analysis}
Fine-tuning refers to further training a language model to directly predict properties of labeled data points using supervised learning\cite{https://doi.org/10.48550/arxiv.1801.06146}. Fine-tuning boosts the model’s performance on a downstream task in part by transferring broader knowledge learned during masked language modelling. The embeddings from the language model are not extracted at any point during fine-tuning. Instead, all the model’s parameters are optimized to classify labeled training data. Both LazBF-ESM and LazDEF-ESM were fine-tuned using supervised learning on their respective data sets. For each model, the same sequences used for masked language modeling were used as the training set for fine-tuning. The same held-out data sets containing sequences unseen during masked language modeling and fine-tuning were used to evaluate the fine-tuned models. Vanilla-ESM was also fine-tuned to classify substrates of the entire lactazole biosynthetic pathway. The accuracy on each held-out data set was calculated for each of the 3 fine-tuned models.

Integrated gradients are an interpretable machine learning technique used to quantify the positive or negative contribution of input features to a model's prediction for a given data point\cite{pmlr-v70-sundararajan17a}. In the context of predicting whether a peptide is the substrate of an enzyme, a positive value for a given residue implies that the residue is important for substrate fitness. A negative value for a given residue suggests that the residue is associated with being a non-substrate. The fine-tuned LazBF model and the fine-tuned LazDEF model were separately used to calculate the integrated gradients for each peptide in the held-out LazBF data set. For each model, and for each residue type, all contributions of that residue across all 50,000 sequences were summed and then divided by the frequency of that residue in the held-out LazBF data, producing two matrices of shape $1 \times 20$ representing the average contribution of each residue type according to the integrated gradients of each model. A similar procedure produced two $1 \times 11$ matrices, representing the average contribution of each position for each model. Finally, a similar procedure produced two $20 \times 11$ matrices, representing the average contribution of each residue type in each position for each model. 

ESM-2 employs a multi-head self-attention mechanism, where each of the 12 layers produce 20 attention heads (240 attention heads in total)\cite{Lin2022}. Each attention head is a 2D matrix $\alpha$ of shape $L \times L$, where $L$ is the length of the tokenized input sequence. The tokenized input sequence includes a ``beginning of sequence" ([BOS]) and an ``end of sequence" ([EOS]) character in addition to the amino acids. Individual attention weights $\alpha_{i, j}$ quantify how much the residue at position $i$ affects the model's representation of the residue at position $j$, with greater values suggesting greater influence. Attention weights have been shown to highlight biological features of proteins including residue-residue contacts and binding sites\cite{https://doi.org/10.48550/arxiv.2006.15222}. The pairwise nature of the self-attention 
mechanism resembles epistatic interactions in protein/peptide fitness landscapes\cite{Starr2016}. Vinogradov \textit{et al.} calculated pairwise epi-scores that attempted to quantify how the fitness of a residue at a given position is affected by residues at other positions \cite{Vinogradov2022}. Thus, we looked for similarities between self-attention matrices and the pairwise epi-scores calculated in previous work for one LazBF and one LazDEF substrate. All 240 attention matrices were obtained for both peptides. 

\section{Results and discussion}

\subsection{Vanilla-ESM Baseline}
We first evaluated the performance of downstream LazBF and LazDEF substrate classification models trained on embeddings from a baseline protein language model (Vanilla-ESM). The performance of each model type was evaluated separately under a high-N, medium-N, and low-N condition defined by the number of sequences used for training. The results of each model type trained on embeddings from Vanilla-ESM -- without any additional masked language modeling -- are displayed in Table 1. Embeddings from Vanilla-ESM perform reasonably well on RiPP biosynthetic enzyme substrate classification tasks, particularly in the high-N condition for the LazBF substrate prediction task. The reasonable performance of Vanilla-ESM embeddings underscores the richness of protein language model representations, which can effectively generalize to novel tasks. Models trained on fewer training samples (i.e., medium-N and low-N) had lower performance. This reflects the importance of having sufficiently large and diverse training data in supervised learning paradigms.

\begin{table}[ht]
  \caption{Classification Accuracy with Vanilla-ESM Embeddings}
  \label{tbl:example}
  \centering
  \resizebox{\textwidth}{!}{
  \begin{tabular}{llcccccc}
    \hline
    & Training Size & SVC & MLP & LR & RF & AB & KNN  \\
    \hline
    LazBF & High-N & \boldmath $90.9\pm0.13$ & $90.2\pm0.31$ & $89.6\pm0.27$ & $88.4\pm0.26$ & $88.8\pm0.42$ & $87.4\pm0.29$  \\
    LazDEF & High-N & \boldmath $83.0\pm0.27$ & $81.5\pm0.19$ & $80.5\pm0.21$ & $79.7\pm0.32$ & $79.4\pm0.48$ & $78.0\pm0.26$  \\
    \hline

    \hline
    & Training Size & SVC & MLP & LR & RF & AB & KNN  \\
    \hline
    LazBF & Medium-N & $88.1\pm0.57$ & $88.0\pm0.26$ & \boldmath $88.2\pm0.09$ & $84.9\pm0.33$ & $85.4\pm0.45$ & $83.8\pm0.80$  \\
    LazDEF & Medium-N & $78.1\pm0.41$ & $76.8\pm0.30$ & \boldmath $79.0\pm0.47$ & $76.4\pm0.51$ & $75.4\pm0.35$ & $73.9\pm0.57$  \\
    \hline

    \hline
    & Training Size & SVC & MLP & LR & RF & AB & KNN  \\
    \hline
    LazBF & Low-N & \boldmath $82.2\pm0.92$ & $81.3\pm0.93$ & $80.9\pm2.07$ & $77.4\pm1.75$ & $76.1\pm1.77$ & $77.0\pm0.80$  \\
    LazDEF & Low-N & $70.8\pm0.97$ & $72.3\pm0.74$ & \boldmath $72.7\pm0.94$ & $69.1\pm1.15$ & $64.8\pm1.35$ & $67.6\pm0.86$  \\
    \hline
  \end{tabular}}
  \begin{tablenotes}
      \item
      Accuracy of support vector classifier (SVC), multi-layer perceptron (MLP), logistic regression (LR), random forest (RF), AdaBoost (AB), and \textit{k}-nearest neighbors classifier (KNN) models trained on embeddings from Vanilla-ESM on both substrate classification tasks. Values are Mean $\pm$ SD. The best performing model in each row is highlighted.
  \end{tablenotes}
\end{table}

\subsection{Masked Language Modeling Improves LazDEF Substrate Classification Performance}
\begin{figure}[!ht]
     \centering
     \hspace{-52pt}
     \includegraphics[scale=0.67]{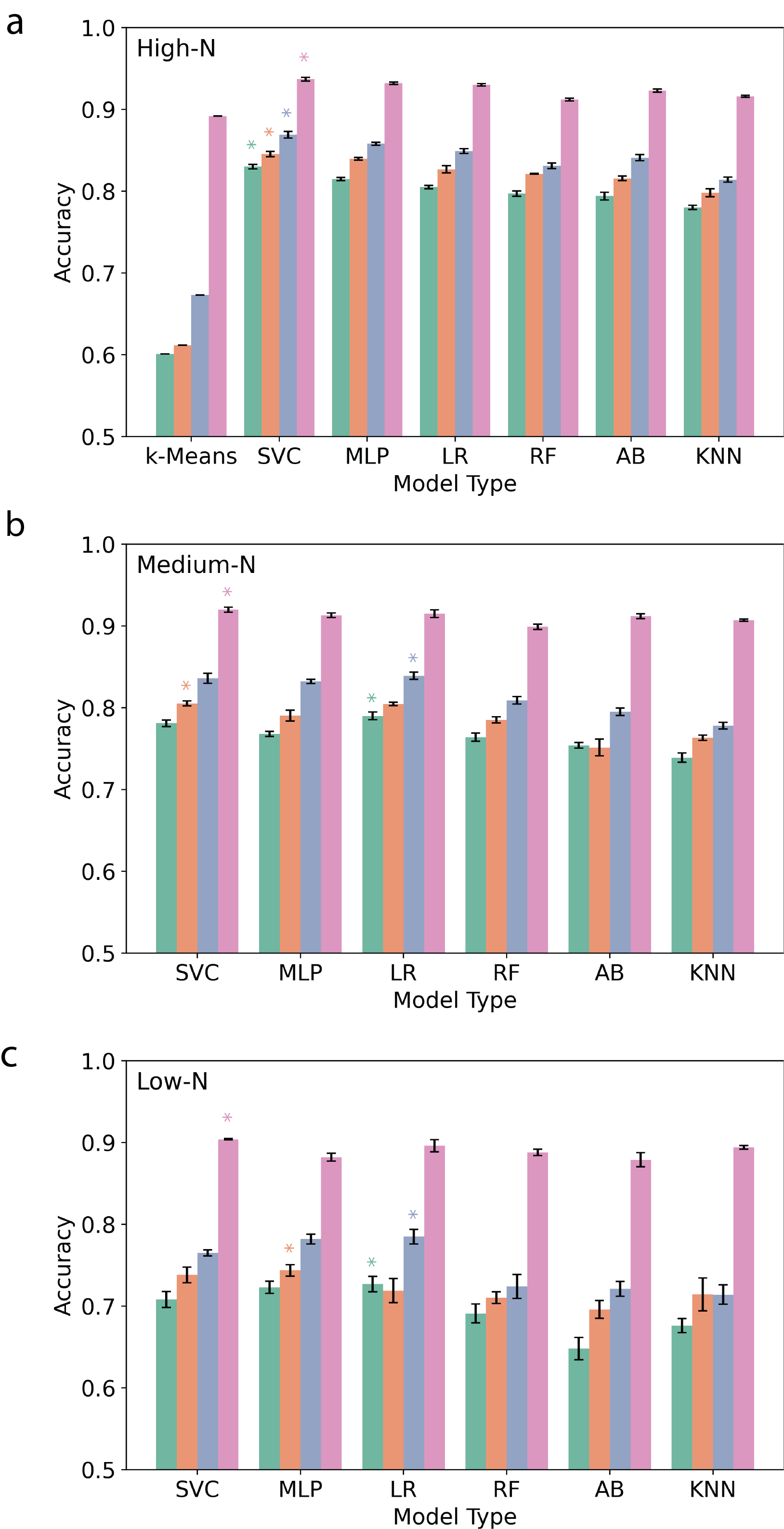}
     \caption{Accuracy of LazDEF substrate classification models trained on embeddings
from Vanilla-ESM (green), ESM trained on a subset of PeptideAtlas (orange), ESM trained on LazBF substrates/non-substrates (blue), and ESM trained on LazDEF substrates/non-substrates (pink) in the a) high-N condition, b) medium-N condition, and c) low-N condition. A star indicates the top performing model for each set of embeddings.
}
     \label{fig:figure4}
\end{figure}

The accuracy of each supervised model type trained on LazDEF substrate/non-substrate embeddings from each of the four language models are presented in Figure \ref{fig:figure4}. The precision, recall, AUROC, and F1 score of each model type is available in Figures S1-S4. LazDEF-ESM produced embeddings that significantly increased the performance of all downstream LazDEF substrate classification models across all training sizes. We suspect that during masked language modeling, the model became attuned to specific features of the LazDEF data set, including the features that distinguish substrates from non-substrates. The model's representations were updated in accordance with these features, allowing for improved discrimination of substrate and non-substrate sequences. 

Strikingly, LazBF-ESM also produced embeddings that significantly increased the performance of LazDEF substrate classification models. Every LazDEF substrate classification model across all training sizes showed a sizable improvement in performance when trained on embeddings from LazBF-ESM, demonstrating that transfer learning improved the performance of the models. Embeddings from Peptide-ESM also improved LazDEF substrate classification models in nearly all cases, but to a lesser extent than LazBF-ESM or LazDEF-ESM embeddings. This indicated that large data sets characterizing the substrate preferences of specific RiPP biosynthetic enzymes provided the most utility in improving RiPP biosynthetic enzyme substrate classification models.

\begin{figure}[t]
     \centering
     \includegraphics[width=\textwidth]{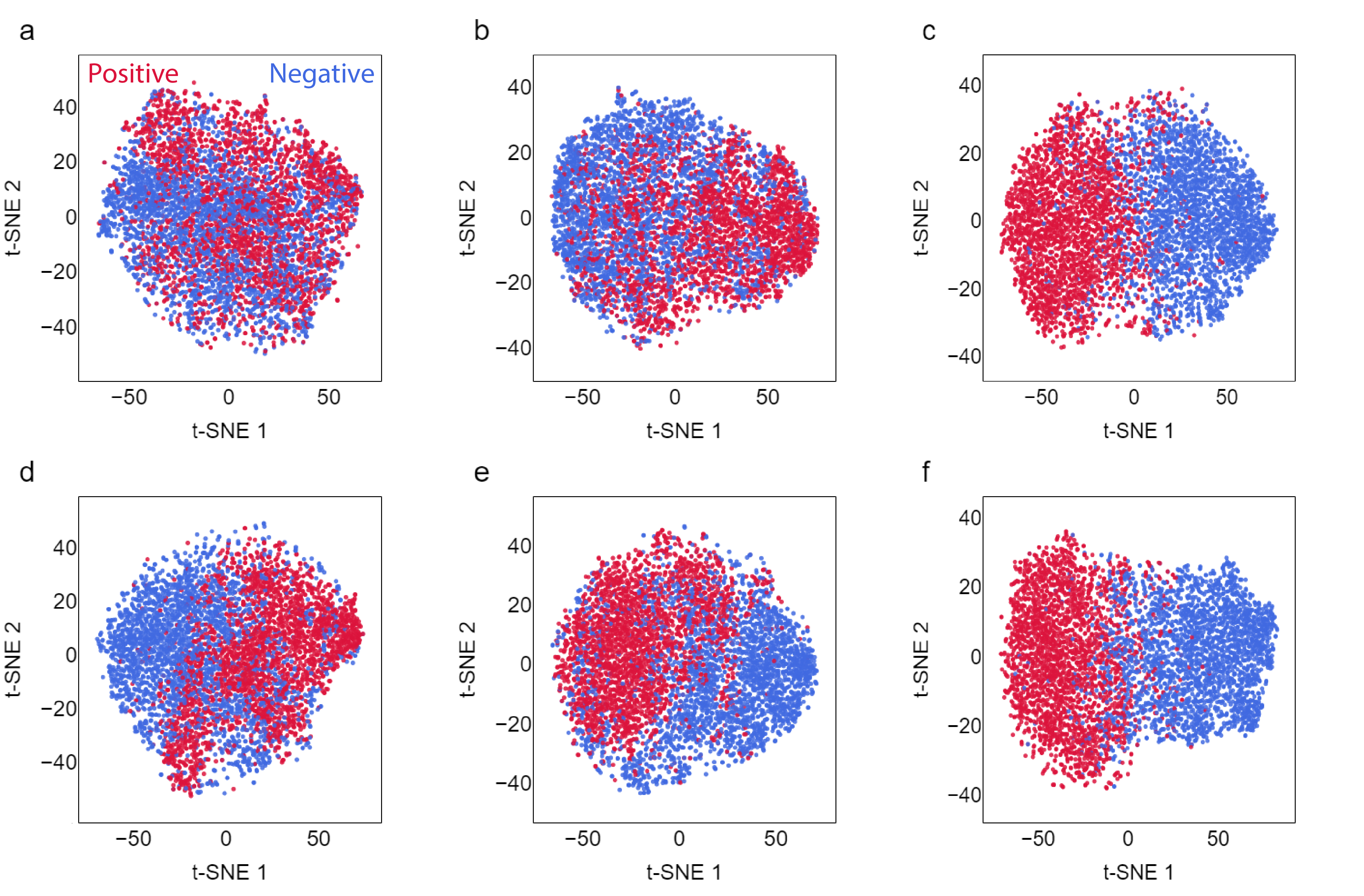}
     \caption{t-SNE visualization of the LazDEF embedding space for a) Vanilla-ESM,
b) ESM trained on LazBF substrates/non-substrates, and c) ESM trained on LazDEF
substrates/non-substrates. t-SNE visualization of the LazBF embedding space for d) Vanilla-
ESM, e) ESM trained on LazDEF substrates/non-substrates, and f) ESM trained on LazBF
substrates/non-substrates. Substrates are red and non-substrates samples are blue.}
     \label{fig:figure5}
\end{figure}

t-SNE was then used to reduce each set of embeddings to two dimensions for visualization. t-SNE plots of the Vanilla-ESM and Peptide-ESM embedding spaces of LazDEF substrates and non-substrates do not show any apparent distinction between substrates and non-substrates (Figures \ref{fig:figure5}, S9). Notably, the LazBF-ESM embedding space shows a visibly higher degree of clustering within substrates and non-substrates than the Vanilla-ESM embedding space. This agrees with the increase in downstream LazDEF substrate classification model performance observed after masked language modeling of the LazBF data set. Finally, the LazDEF-ESM embedding space shows the most obvious segregation (Figure \ref{fig:figure5}). The increased ability to distinguish LazDEF substrates/non-substrates suggests that using embeddings from a language model trained on a large data set relevant to the task of interest can greatly increase the predictive power of downstream classifiers through transfer learning.

\subsection{Masked Language Modeling of Either Data Set Improves LazBF Substrate Classification Performance}
\begin{figure}[!ht]
     \centering
     \hspace{-52pt}
     \includegraphics[scale=0.675]{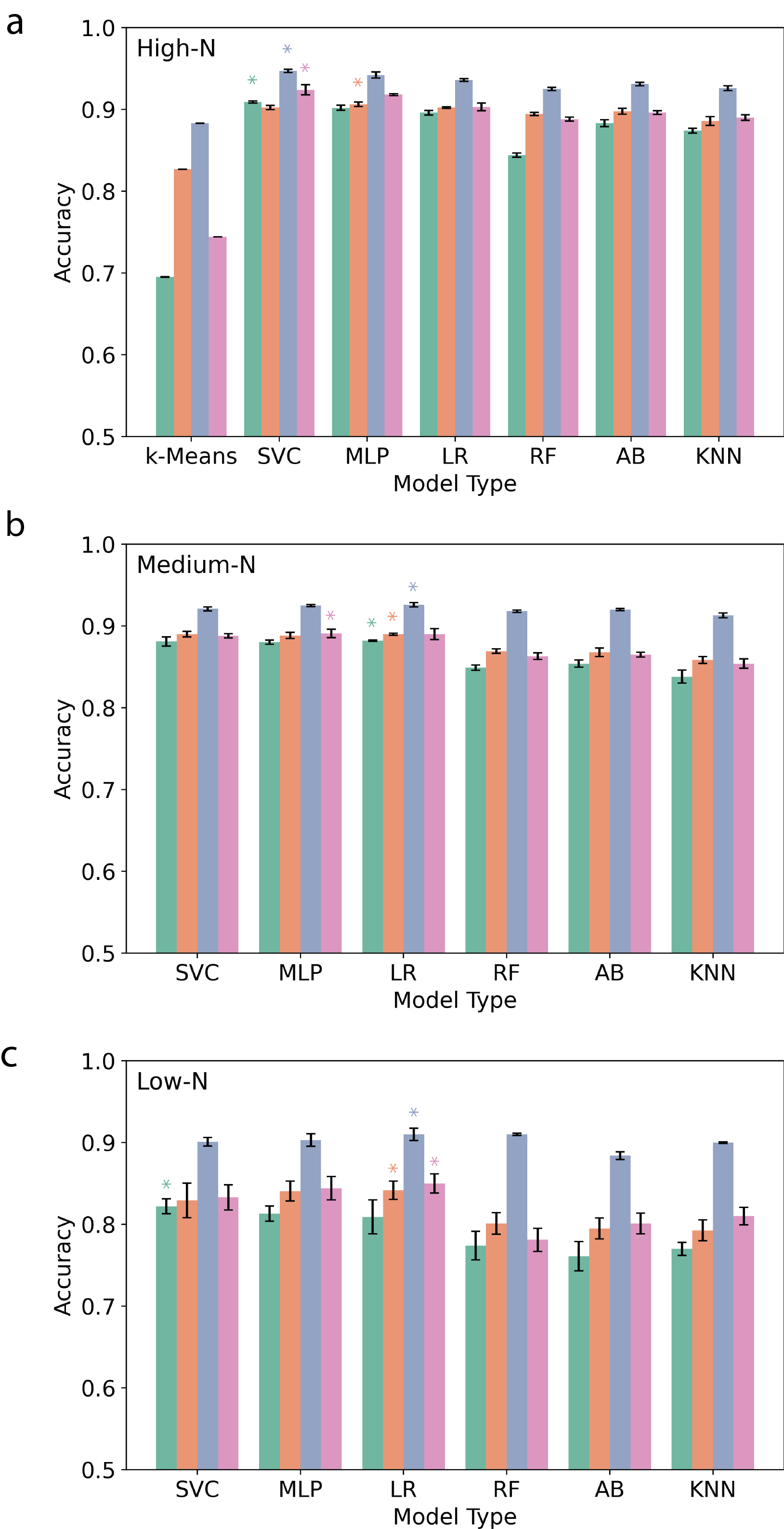}
     \caption{Accuracy of LazBF substrate classification models trained on embeddings from Vanilla-ESM (green), ESM trained on a subset of PeptideAtlas (orange), ESM trained on LazBF substrates/non-substrates (blue), and ESM trained on LazDEF substrates/non-substrates (pink) in the a) high-N condition, b) medium-N condition, and c) low-N condition. A star indicates the top performing model for each set of embeddings.}
     \label{fig:figure6}
\end{figure}

The accuracy of each model type trained on embeddings of LazBF substrates/non-substrates from each of the 4 protein language models are presented in Figure \ref{fig:figure6}. The precision, recall, AUROC, and F1 score of each model type is available in Figures S5-S8. Similarly, LazBF-ESM produced embeddings that significantly improved the performance of both unsupervised \textit{k}-Means clustering and supervised classification models of LazBF substrates across all training sizes. LazDEF-ESM also produced embeddings that improved the performance of most LazBF substrate classification models. In the high-N condition, all models showed performance increases, with unsupervised \textit{k}-Means clustering showing the most improvement. Most supervised models trained using the medium-N and low-N conditions also showed improved performance. SVC and MLP showed the largest and most consistent increases across these two conditions. Expectedly, the low-N condition produced models with higher variance, which likely contributed to more unstable results. In most cases, LazDEF-ESM embeddings also outperformed Peptide-ESM embeddings. 

A t-SNE plot of the LazBF substrate/non-substrate embeddings from Vanilla-ESM and Peptide-ESM show an already apparent distinction between substrates and non-substrates (Figures \ref{fig:figure5}, S9). This suggests that the pretrained model is sensitive to differences inherent in LazBF substrates and non-substrates. The visual divergence of substrates and non-substrates is arguably more apparent in the embedding space of LazDEF-ESM (Figure \ref{fig:figure5}e). Predictably, the embedding space of LazBF-ESM shows the most dramatic separation of substrates from non-substrates (Figure \ref{fig:figure5}f). This is consistent with large increases in downstream LazBF substrate classification model performance after masked language modeling of the LazBF data set.

The observation that LazBF substrate classifiers showed improved performance when trained on embeddings from LazDEF-ESM suggests that information relevant to LazBF classification was learned during masked language modeling of the LazDEF substrates/non-substrates. However, Vanilla-ESM embeddings already showed good performance on LazBF prediction tasks. We suspect that this left less room for improvement through masked language modeling of the LazDEF data set. However, any improvement is compelling given that 1) LazBF and LazDEF catalyze disparate transformations and 2) the substrate fitness landscapes of LazBF and LazDEF are reported to be divergent from one another, particularly in the degree to which pairwise positional epistasis affects fitness\cite{Vinogradov2022}. Tanimoto similarity is a common metric used to quantify the chemical similarity between small molecules and peptides. The average Tanimoto similarity between peptides in the held-out LazBF and held-out LazDEF substrate data sets was calculated to be $0.354\pm0.031$, suggesting that the data sets contained relatively dissimilar sequences. The results of this and the previous section show that knowledge learned during the unsupervised modeling of RiPP biosynthetic enzyme substrates/non-substrates can be transferred to other tasks, particularly those that involve related but distinct enzymes from the same biosynthetic pathway. Additionally, unsupervised modeling of RiPP biosynthetic enzyme substrates/non-substrates appear to produce better representations than unsupervised modeling of diverse peptides. 

Despite catalyzing different transformations, both LazBF and LazDEF bind LazA precursor peptides as substrates. Therefore, there is expected to be some degree of similarity between the substrate preferences of the two enzymes. However, we observed that more information about LazDEF substrate preferences was learned from masked language modeling of LazBF substrate preferences than vice versa. We hypothesize that this asymmetry results from the order of the post-translational modifications that occur during lactazole biosynthesis. In nature, LazDEF modifies LazA precursor peptides prior to LazBF\cite{Vinogradov2020c}. Therefore, self-supervised modeling of LazBF substrate preferences learns the biophysical features of substrates that are likely to have been modified by LazDEF. However, the opposite is not necessarily true. This presents an intuitive explanation as to why transfer learning showed greater success at improving LazDEF substrate classification models.


\subsection{Fine-Tuned Language Model Performance on RiPP Biosynthetic Enzyme Classification Tasks} 

LazBF-ESM and LazDEF-ESM were then trained to classify the substrates of their respective data sets through a training procedure called fine-tuning. Vanilla-ESM was also fine-tuned to classify substrates of the entire lactazole biosynthetic pathway. Each fine-tuned model showed excellent performance on its respective held-out data set ($>$0.95 accuracy in each case). We also evaluated how well each fine-tuned model performed on the other held-out data sets without any further training (Table 2). The fine-tuned LazBF-ESM model showed no ability to classify LazDEF substrates, and showed little ability to classify substrates for the entire pathway after supervised training. In contrast, the fine-tuned LazDEF model achieved 0.697 accuracy  on the held-out LazBF substrate data set, likely due in part to the LazBF data set being more enriched (Figure \ref{fig:figure5}d). This model also showed some ability to classify substrates of the entire lactazole biosynthetic pathway. Finally, the supervised model trained to classify substrates of the entire pathway showed some ability to classify LazBF and LazDEF substrates without any further training. 

\begin{figure}[ht]
     \includegraphics[width=\textwidth]{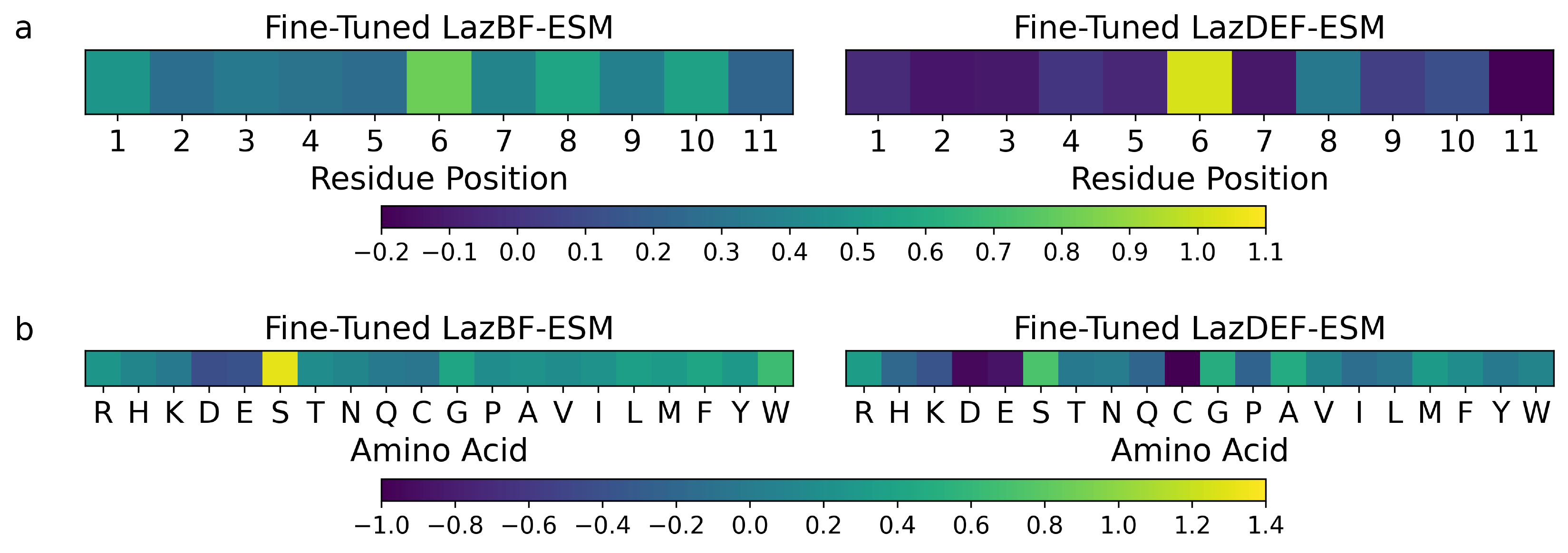}
     \caption{Fine-tuned LazBF-ESM and fine-tuned LazDEF-ESM produce correlated integrated gradients for LazBF substrates/non-substrates. a) The average contribution of each position to substrate fitness shows a 0.81 spearmanr between the two models. b) The average contribution of each amino acid to substrate fitness shows a 0.80 spearmanr between the two models.}
     \label{fig:figure7}
\end{figure}

Integrated gradients can quantify how individual residues contribute to a model’s prediction. Inspired by the performance of LazDEF-ESM on the LazBF substrate classification task, we looked for similarities between the integrated gradients for LazBF substrates/non-substrates from both models (Figure \ref{fig:figure7}). We observed that the average contribution of each residue type from fine-tuned LazDEF-ESM strongly correlated with the average contribution of each residue type from fine-tuned LazBF-ESM, with a spearman coefficient of 0.80 (Figure \ref{fig:figure7}a). Similarly, the average contribution of each position from both fine-tuned models showed a 0.81 spearman coefficient (Figure \ref{fig:figure7}b). The average contribution of each residue type in each position also showed a moderate correlation (0.59 spearman coefficient). These correlations exist despite fine-tuned LazDEF-ESM having never been trained on LazBF substrates. Therefore, to some extent, fine-tuned RiPP biosynthetic enzyme prediction models can produce valid and interpretable predictions about distinct, but related prediction tasks.

\begin{table}
\caption{Classification Accuracy Fine-Tuned Models}
  \label{tbl:example}
  \begin{tabular}{lllllllll}
    \hline
    & Supervised LazBF & Supervised LazDEF & Supervised LazBCDEF  \\
    \hline
    LazBF test set  & $99.3\%$ & $69.7\%$ & $64.8\%$ \\
    LazDEF test set & $50.9\%$ & $99.2\%$ & $58.8\%$ \\
    LazBCDEF test set & $52.3\%$ & $64.1\%$ & $95.9\%$ \\
    \hline
  \end{tabular}
\end{table}

\subsection{Attention Analysis}

\begin{figure}[ht]
     \includegraphics[width=0.85\textwidth]{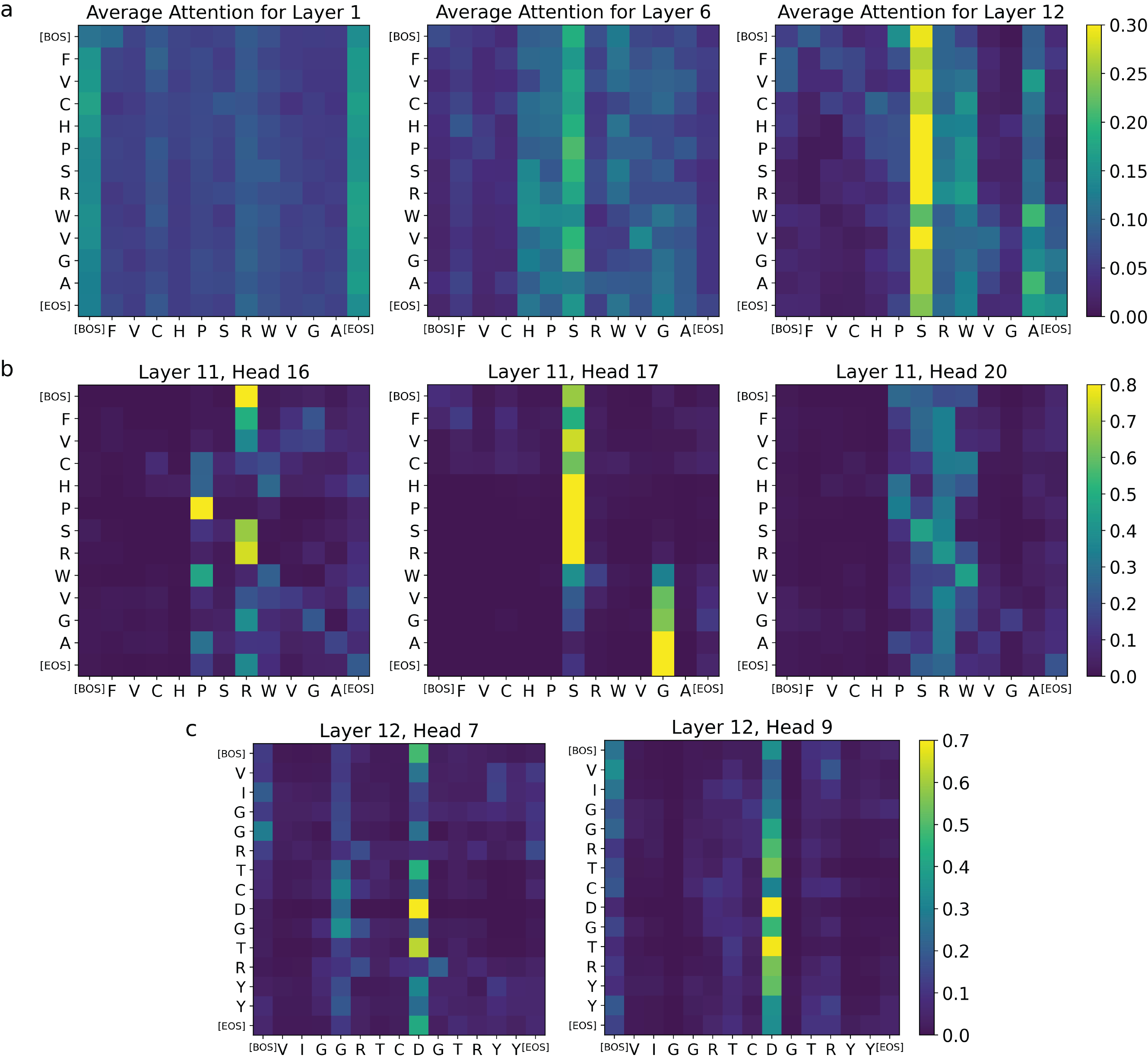}
     \caption{Attention maps from the fine-tuned LazBF-ESM. [BOS] and [EOS] tokens mark the ``beginning of sequence” and ``end of sequence" respectively. a) Middle and later layers focus on specific residues and motifs. b) Attention heads from the penultimate layer highlight a motif with high pairwise epi-scores in a LazBF substrate. c) Attention heads from the final layer highlight a residue important for substrate fitness in a LazDEF substrate.}
     \label{fig:figure8}
\end{figure}

Attention matrices describe the model's perceived relevance or association between each pair of tokenized residues, including the [BOS] and [EOS] tokens added to the beginning and the end of the peptide respectively (see methods). Higher values between a pair of tokens indicates greater relevance between them. Analyzing attention matrices can provide insight into which residues the model regards as important. We observe a general trend in which the attention heads from earlier layers focus mainly on the [BOS] and [EOS] tokens, while heads from later layers dedicate significant attention to specific residues or motifs (Figures \ref{fig:figure8}a, S10). Our observation that the model's attention mechanism `zeros-in' on important residues is consistent with the widespread claim that the per-layer representations of protein language models are hierarchical in nature, with earlier layers encoding low-level features and later layers encoding more global representations of structure and/or function\cite{https://doi.org/10.48550/arxiv.2006.15222}. 



Previous work utilized predictive machine learning models to calculate the pairwise epi-scores for LazBF substrates. Pairwise epi-score values provide an estimate of the strength with which amino acids in the core peptide region affect each other's fitness\cite{Vinogradov2022}. The self-attention mechanism found in transformer models resembles pairwise epi-scores by quantifying the degree to which one amino acid affects the representations of other amino acids in the peptide\cite{Starr2016, https://doi.org/10.48550/arxiv.2006.15222}. For the LazBF substrate FVCHPSRWVGA, the computed pairwise epi-scores suggest that a His4-Pro5-Ser6-Arg7-Trp8 motif contributes to the fitness of the peptide\cite{Vinogradov2022}. Figure \ref{fig:figure8}b shows that multiple attention heads in the 11th layer of the fine-tuned LazBF-ESM dedicate attention between pairs of amino acids within this motif. This suggests that the supervised protein language model's attention mechanism is somewhat able to highlight epistatic interactions and provide a rough idea of which residues are important for fitness. 

Surprisingly, we observe that the fine-tuned version of LazBF-ESM also highlights some epistatic features of the LazDEF substrate VIGGRTCDGTRYY (Figure \ref{fig:figure8}c). Precalculated epi-scores for this peptide indicate that Asp8 has numerous positive and negative epistatic interactions with surrounding peptides including Thr6, Gly9, and Arg11. We find that multiple heads from the last layer of our fine-tuned LazBF-ESM dedicate significant attention between Asp8 and nearby residues, thus highlighting Asp8 as an important residue.

\section{Conclusions}
In this work, we enhanced the performance of protein language model embeddings for RiPP biosynthetic enzyme substrate prediction tasks by performing masked language modeling of substrate/non-substrate data. We applied transfer learning to improve the performance of peptide substrate prediction models for distinct enzymes from the same biosynthetic pathway. A limited number of studies have explored transfer learning in the domain of enzyme substrate prediction, and, to the best of our knowledge, this is the first work to investigate transfer learning between RiPP biosynthetic enzymes. 

We focused on LazBF and LazDEF, a serine dehydratase and azole synthetase respectively, from the lactazole biosynthesis pathway. Masked language modeling was used to train two peptide language models on data sets comprised of LazA sequences labeled as substrates or non-substrates for LazBF and LazDEF respectively. An additional peptide language model was trained on a diverse set of non-LazA peptides. We found that all peptide language models produced embeddings that increased the performance of downstream classification models on both substrate prediction tasks. The LazBF/DEF models provided the largest increases in performance. This suggested some information is shared between the two fitness landscapes, and that masked language modeling of one data set allowed the model to learn important features of the other data set. The performance enhancements were most significant for downstream LazDEF classification models, including the medium-N and low-N conditions. Our workflow enhances the ability to classify RiPP biosynthetic enzyme substrates in limited data regime. This is attractive in the context of peptide engineering, where it could expedite peptide design and discovery by reducing the need for comprehensive experimental profiling. 

We also demonstrated that interpretable machine learning techniques are somewhat transferable between similar RiPP biosynthetic enzyme classification tasks. Specifically, we found that the integrated gradients for LazBF peptides from a supervised LazDEF model correlated with the integrated gradients from a supervised LazBF model. Due to the increasing abundance of sequence data and rapid advances in next-generation sequencing technology, we anticipate the development of large peptide data sets suitable for masked language modeling. Coupled with the growing size and sophistication of protein language models, we expect masked language modeling and transfer learning to aid enzyme substrate prediction tasks especially in cases where large data sets for related enzymes are available. 

\begin{acknowledgement}

The authors thank Dr. Alexander A. Vinogradov, Dr. Yuki Goto, and Dr. Hiroaki Suga for sharing the data sets used in this study. JDC thanks Song Yin for his comments. DS acknowledges support from NIH grants R35GM142745 and R21AI167693.
\end{acknowledgement}

\begin{suppinfo}

Source code for data analysis, downstream model training and validation, along with trained model weights are available at \url{https://github.com/ShuklaGroup/LazBFDEF}.

\end{suppinfo}

\bibliography{LazBFDEF}
\end{document}


    \begin{figure}
        \includegraphics[scale=0.75]{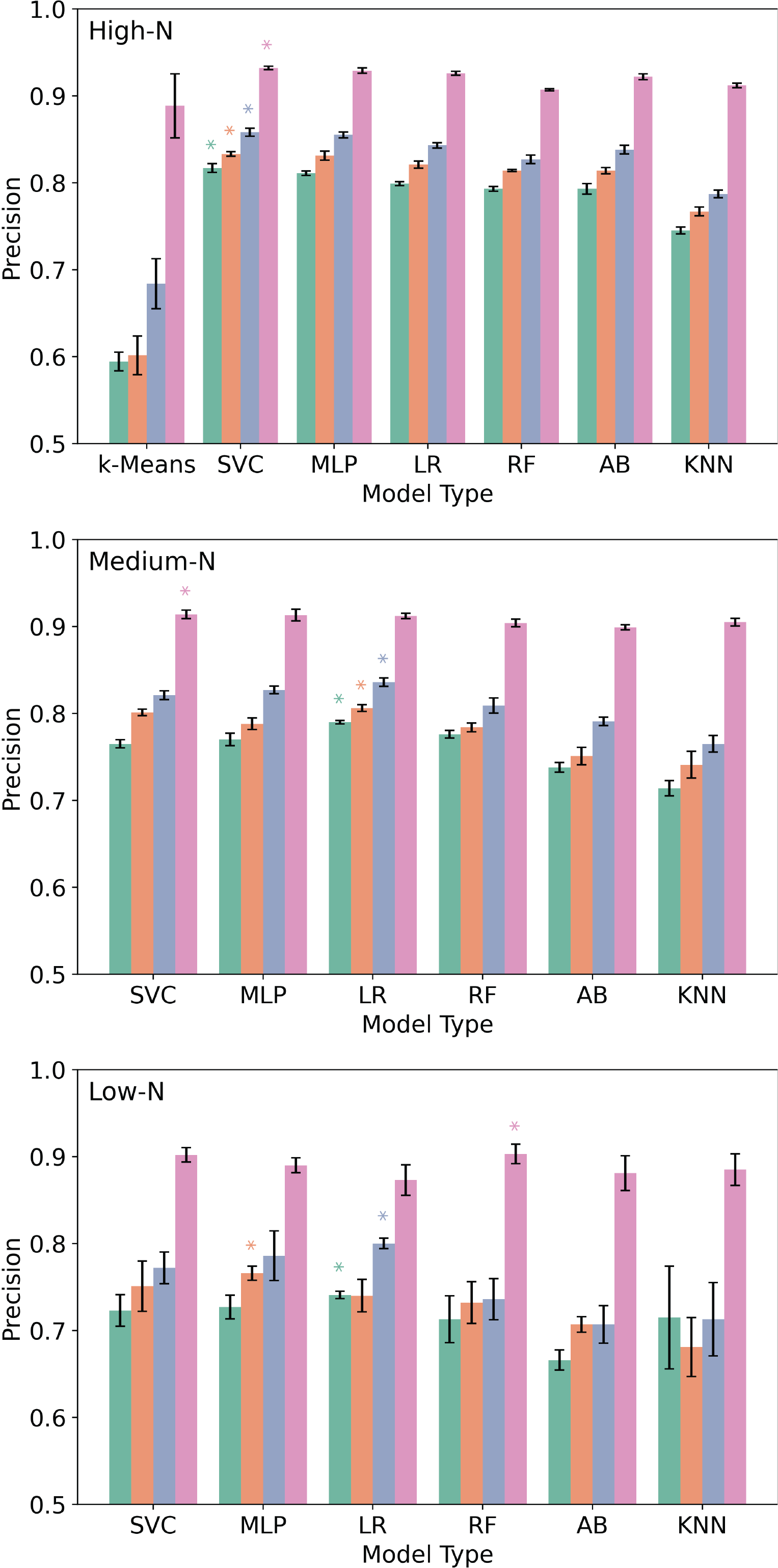}
        \caption{Precision of LazDEF substrate classification models trained on embeddings
from Vanilla-ESM (green), ESM trained on a subset of PeptideAtlas (orange), ESM trained on LazBF substrates/non-substrates (blue), and
ESM trained on LazDEF substrates/non-substrates (pink) in the a) high-N condition, b)
medium-N condition, and c) low-N condition. A star indicates the top performing model for
each set of embeddings.}
        \label{fig:figS1}
    \end{figure}

    \begin{figure}
        \includegraphics[scale=0.75]{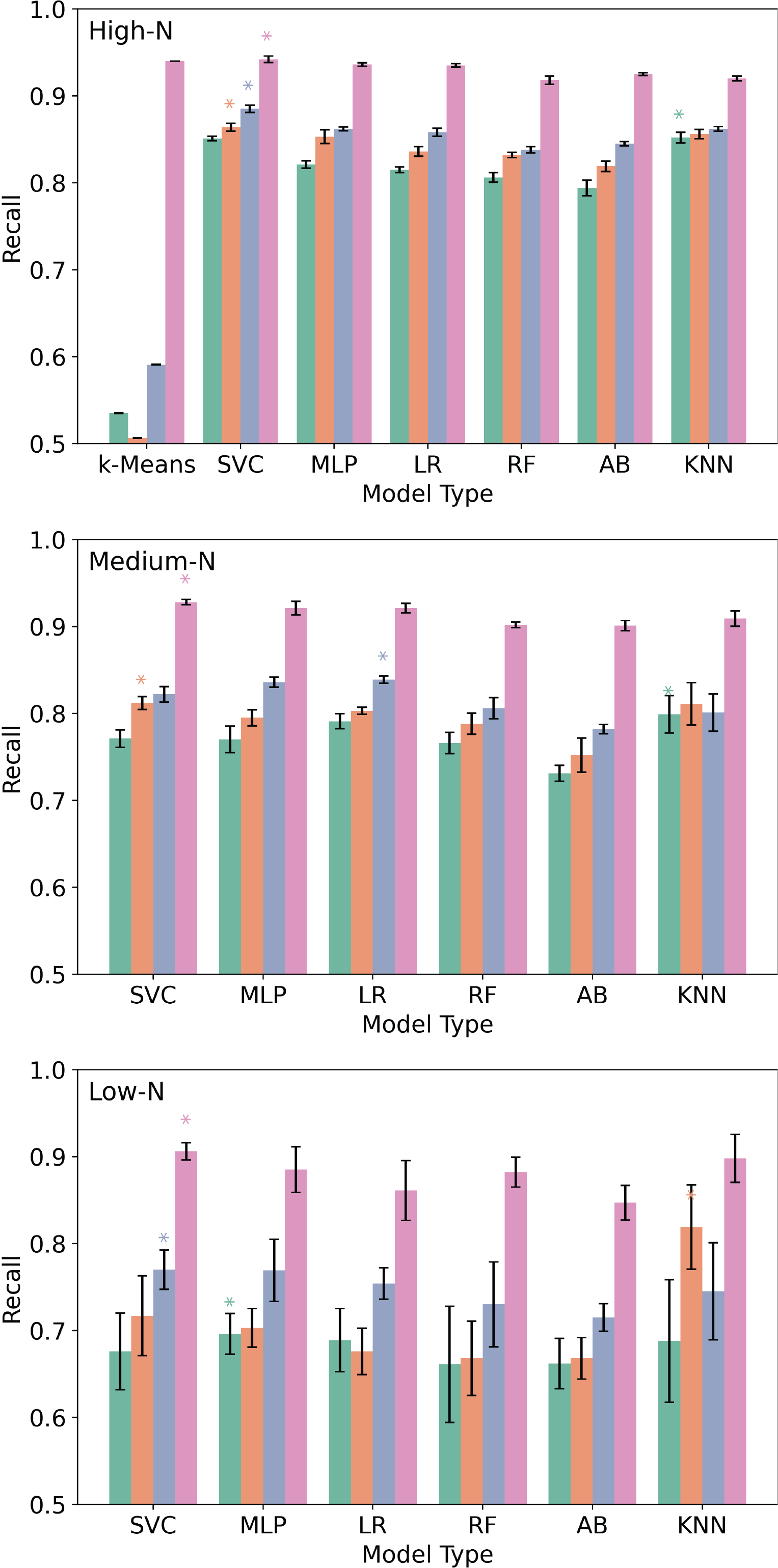}
        \caption{Recall of LazDEF substrate classification models trained on embeddings
from Vanilla-ESM (green), ESM trained on a subset of PeptideAtlas (orange), ESM trained on LazBF substrates/non-substrates (blue), and
ESM trained on LazDEF substrates/non-substrates (pink) in the a) high-N condition, b)
medium-N condition, and c) low-N condition. A star indicates the top performing model for
each set of embeddings.}
        \label{fig:figS2}
    \end{figure}

    \begin{figure}
        \includegraphics[scale=0.75]{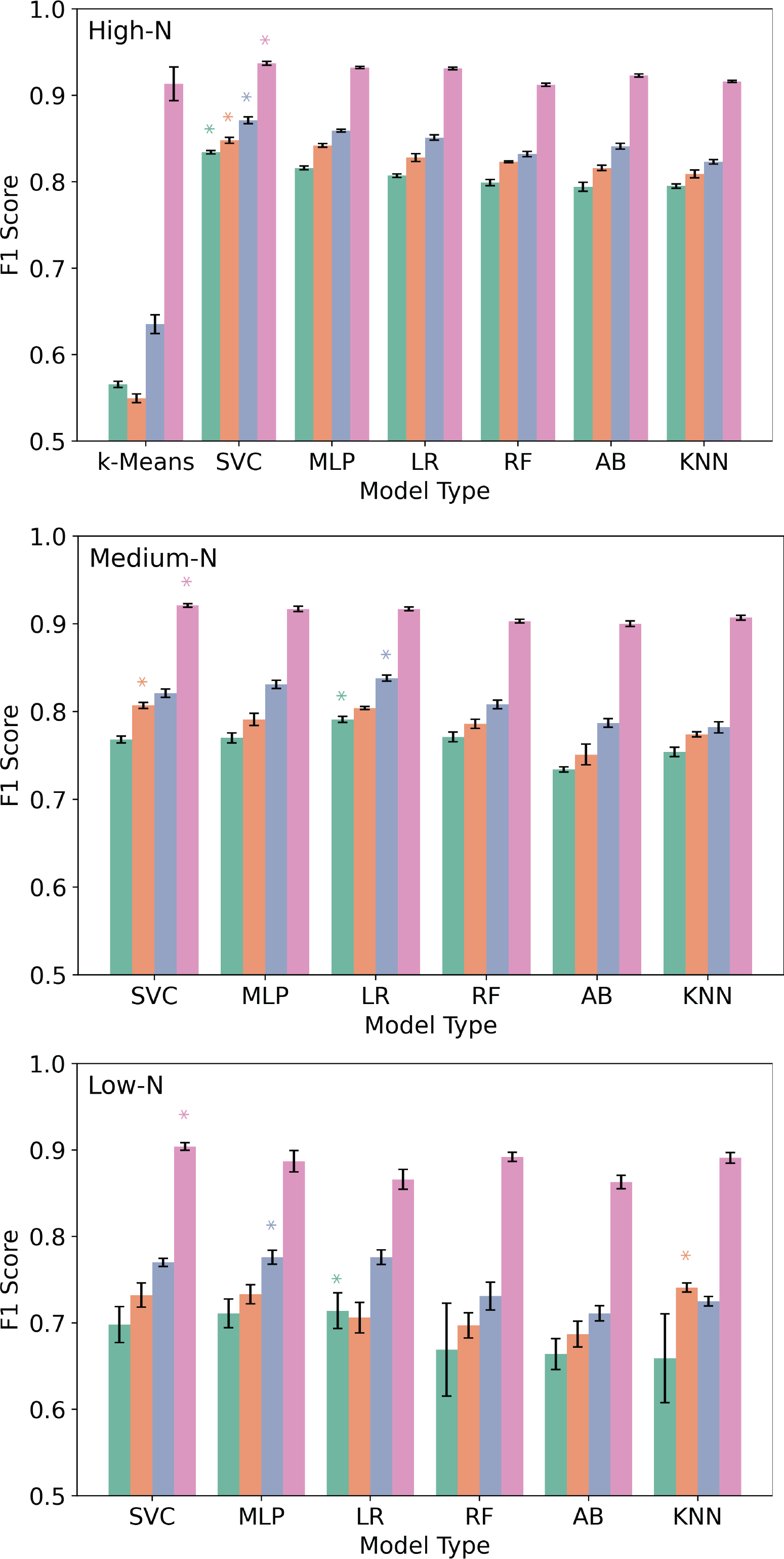}
        \caption{F1 score of LazDEF substrate classification models trained on embeddings
from Vanilla-ESM (green), ESM trained on a subset of PeptideAtlas (orange), ESM trained on LazBF substrates/non-substrates (blue), and
ESM trained on LazDEF substrates/non-substrates (pink) in the a) high-N condition, b)
medium-N condition, and c) low-N condition. A star indicates the top performing model for
each set of embeddings.}
        \label{fig:figS3}
    \end{figure}

    \begin{figure}
        \includegraphics[scale=0.75]{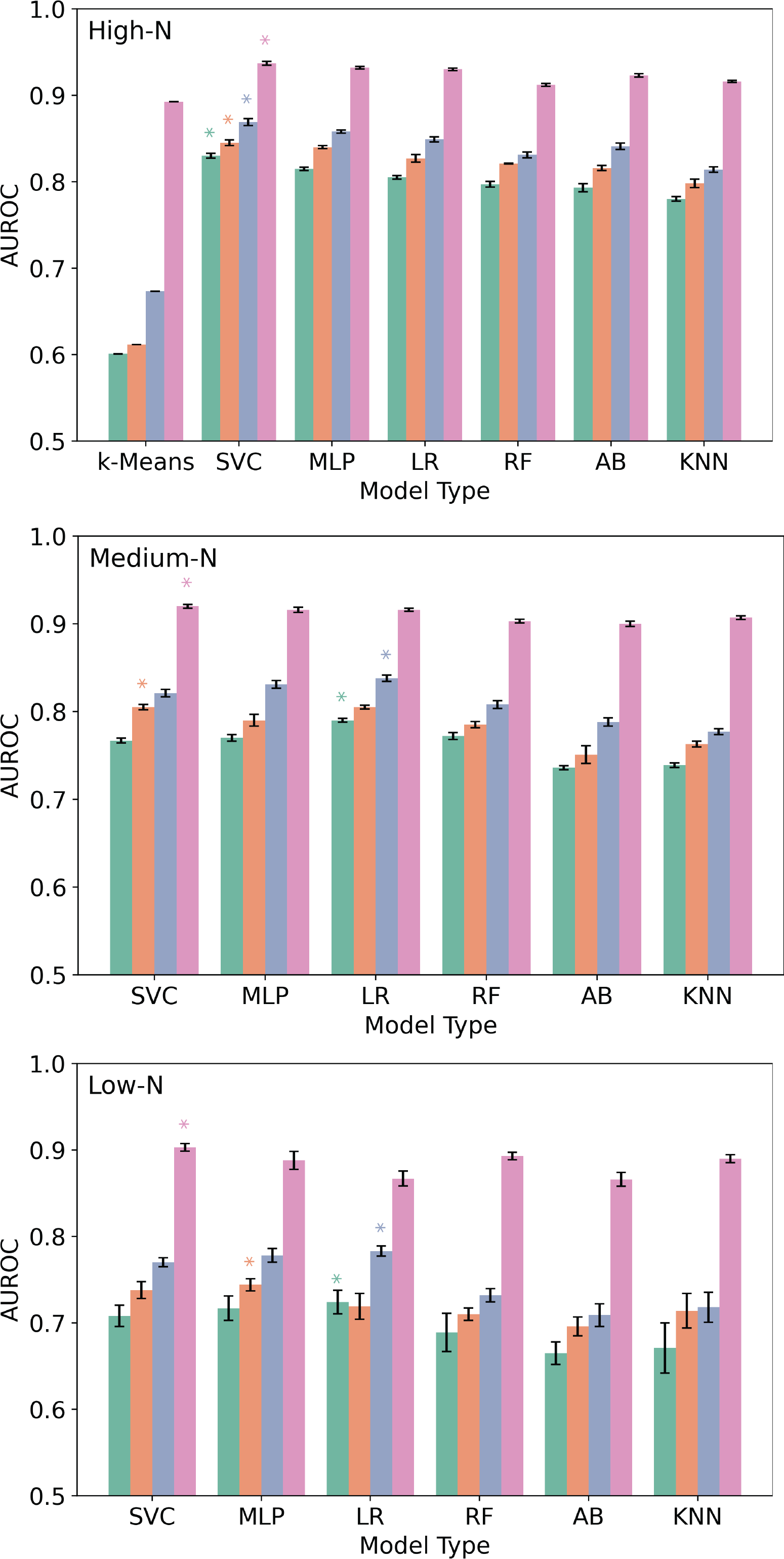}
        \caption{AUROC of LazDEF substrate classification models trained on embeddings
from Vanilla-ESM (green), ESM trained on a subset of PeptideAtlas (orange), ESM trained on LazBF substrates/non-substrates (blue), and
ESM trained on LazDEF substrates/non-substrates (pink) in the a) high-N condition, b)
medium-N condition, and c) low-N condition. A star indicates the top performing model for
each set of embeddings.}
        \label{fig:figS4}
    \end{figure}

    \begin{figure}
        \includegraphics[scale=0.75]{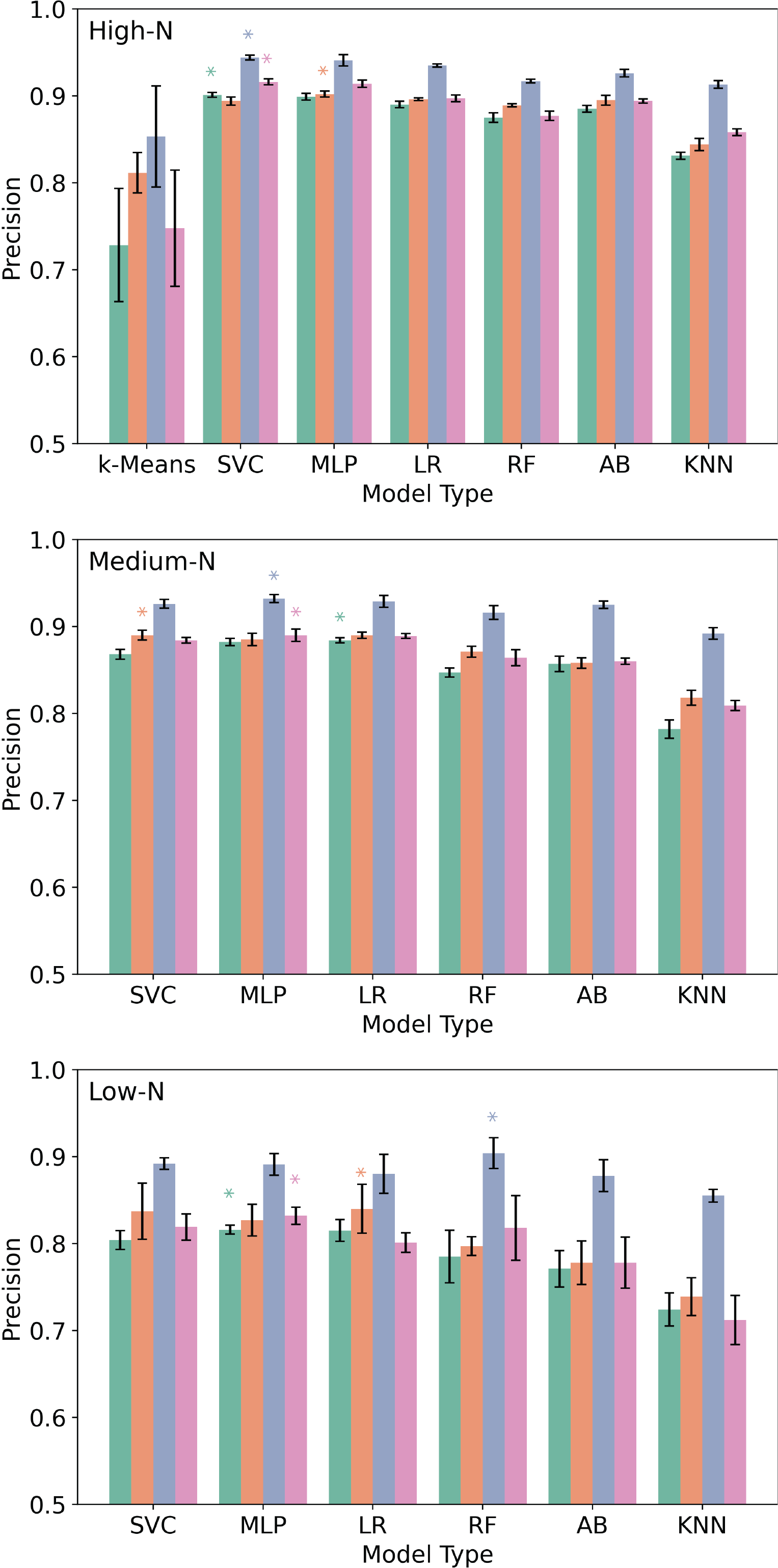}
        \caption{Precision of LazBF substrate classification models trained on embeddings
from Vanilla-ESM (green), ESM trained on a subset of PeptideAtlas (orange), ESM trained on LazBF substrates/non-substrates (blue), and
ESM trained on LazDEF substrates/non-substrates (pink) in the a) high-N condition, b)
medium-N condition, and c) low-N condition. A star indicates the top performing model for
each set of embeddings.}
        \label{fig:figS5}
    \end{figure}

    \begin{figure}
        \includegraphics[scale=0.75]{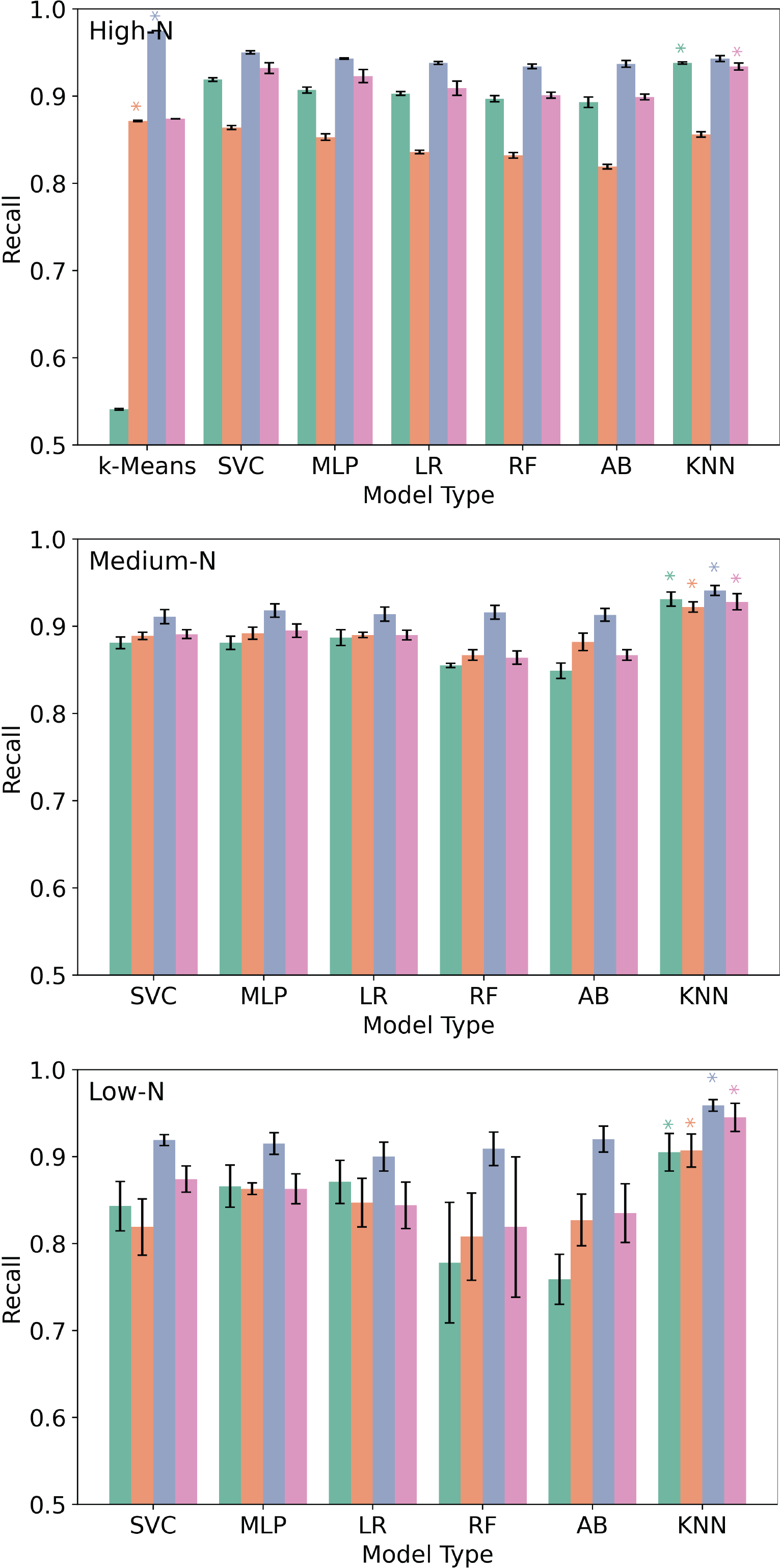}
        \caption{Recall of LazBF substrate classification models trained on embeddings
from Vanilla-ESM (green), ESM trained on a subset of PeptideAtlas (orange), ESM trained on LazBF substrates/non-substrates (blue), and
ESM trained on LazDEF substrates/non-substrates (pink) in the a) high-N condition, b)
medium-N condition, and c) low-N condition. A star indicates the top performing model for
each set of embeddings.}
        \label{fig:figS6}
    \end{figure}

    \begin{figure}
        \includegraphics[scale=0.75]{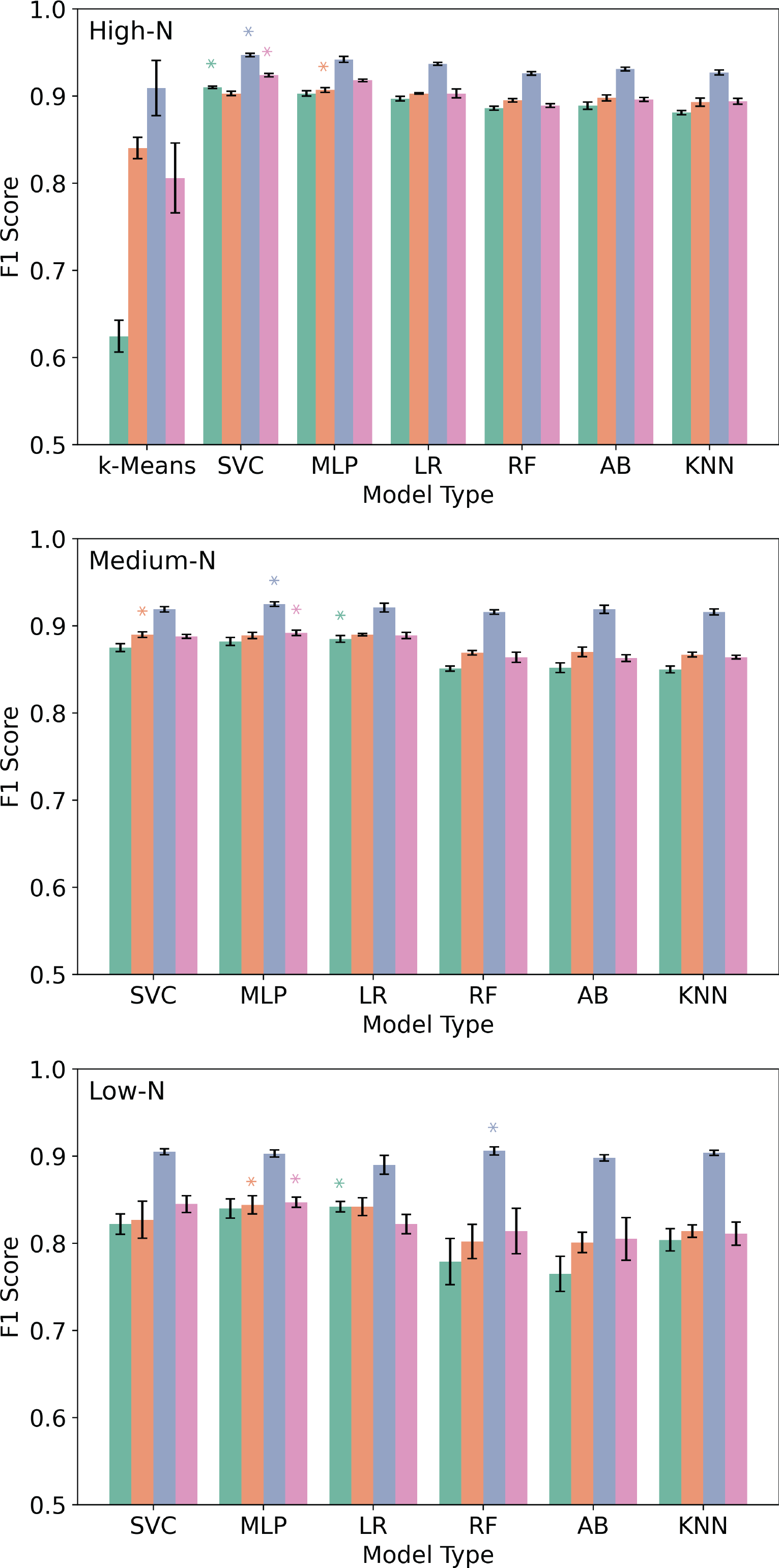}
        \caption{F1 score of LazBF substrate classification models trained on embeddings
from Vanilla-ESM (green), ESM trained on a subset of PeptideAtlas (orange), ESM trained on LazBF substrates/non-substrates (blue), and
ESM trained on LazDEF substrates/non-substrates (pink) in the a) high-N condition, b)
medium-N condition, and c) low-N condition. A star indicates the top performing model for
each set of embeddings.}
        \label{fig:figS7}
    \end{figure}

    \begin{figure}
        \includegraphics[scale=0.75]{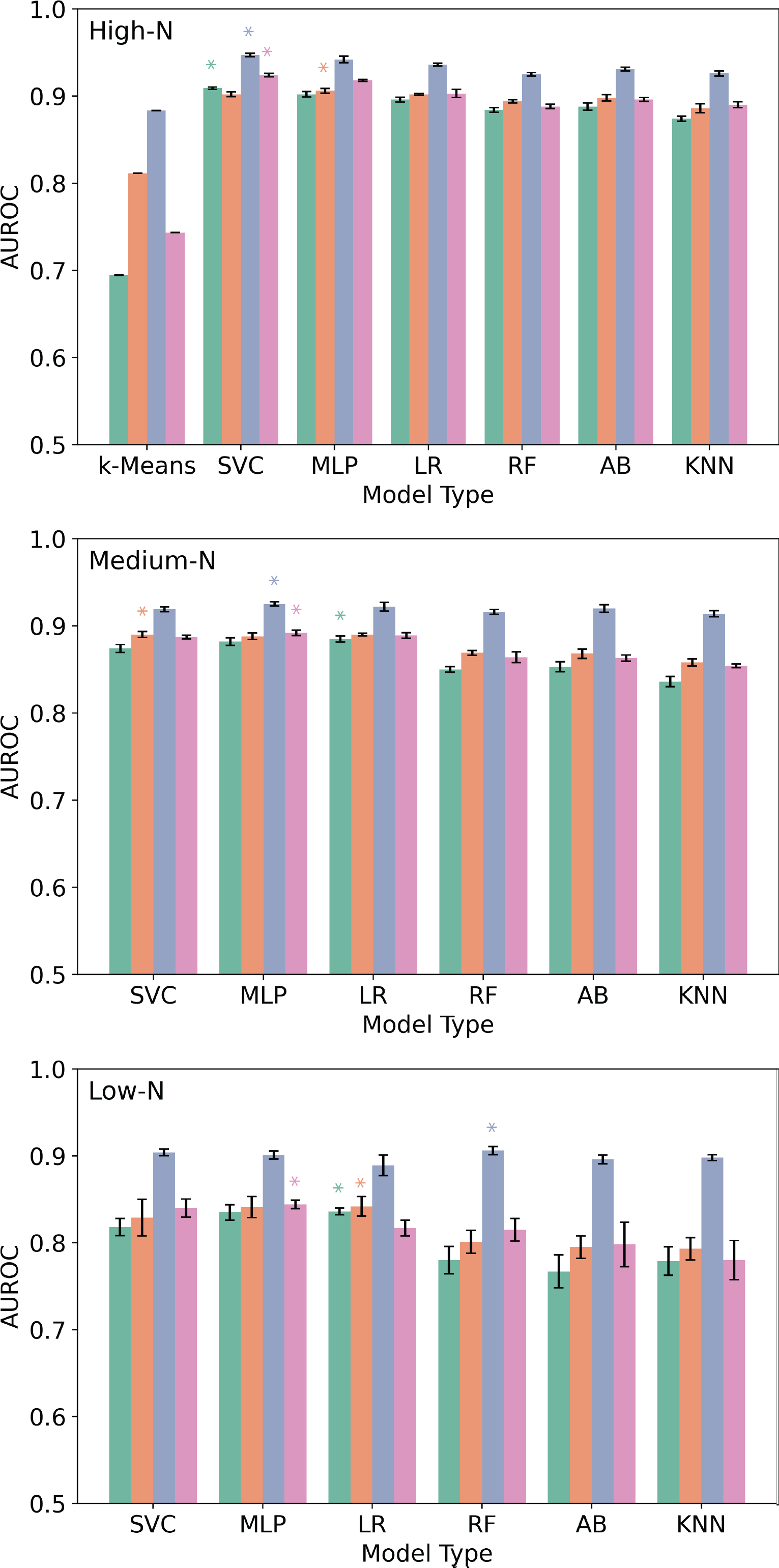}
        \caption{AUROC of LazBF substrate classification models trained on embeddings
from Vanilla-ESM (green), ESM trained on a subset of PeptideAtlas (orange), ESM trained on LazBF substrates/non-substrates (blue), and
ESM trained on LazDEF substrates/non-substrates (pink) in the a) high-N condition, b)
medium-N condition, and c) low-N condition. A star indicates the top performing model for
each set of embeddings.}
        \label{fig:figS8}
    \end{figure}

    \begin{figure}
        \hspace{10pt}
        \includegraphics[width=\textwidth]{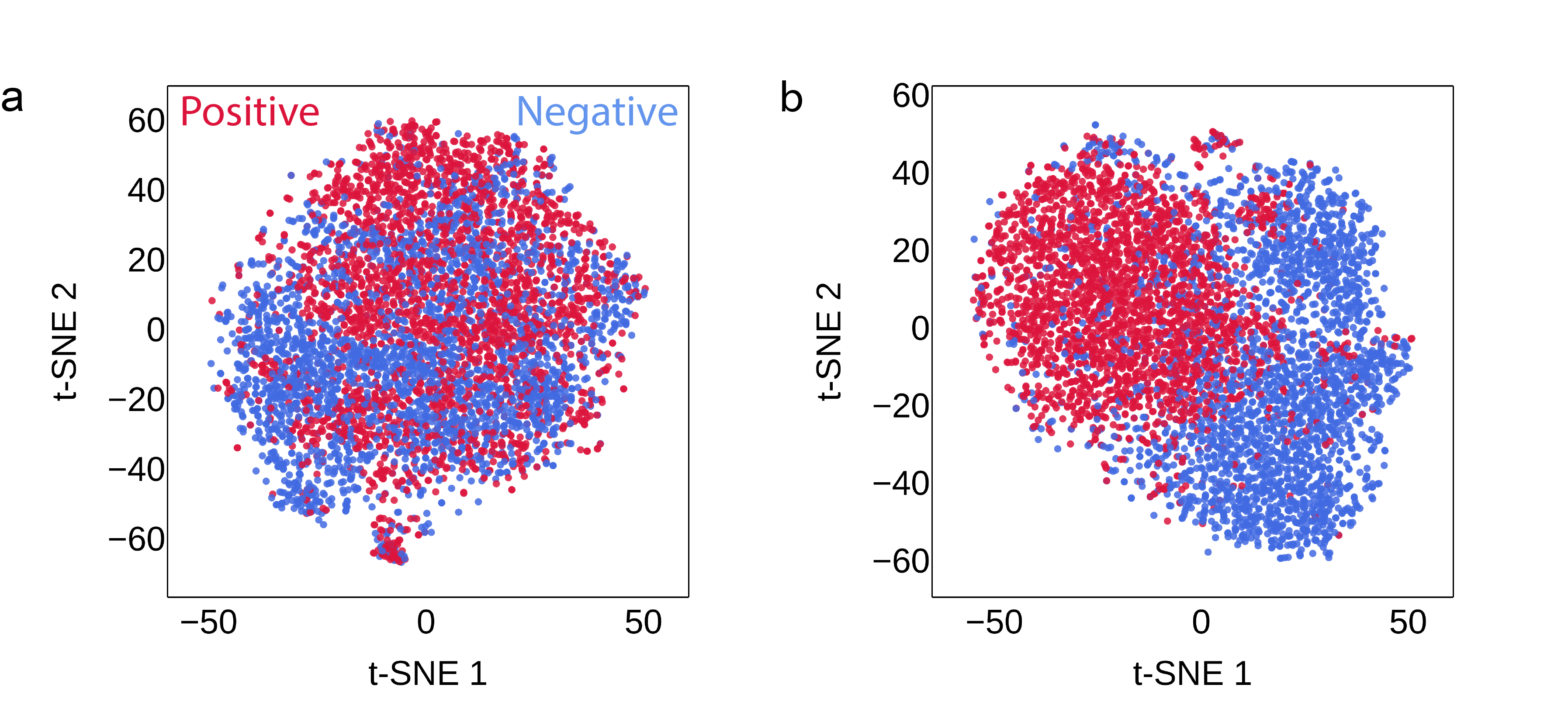}
        \caption{t-SNE visualization of the embedding space from ESM trained on a subset of PeptideAtlas for a) LazDEF substrates/non-substrates, and b) LazBF
substrates/non-substrates. Substrates are red and non-substrates samples are blue.}
        \label{fig:figureS9}
    \end{figure}

    \begin{figure}
    \includegraphics[width=\textwidth]{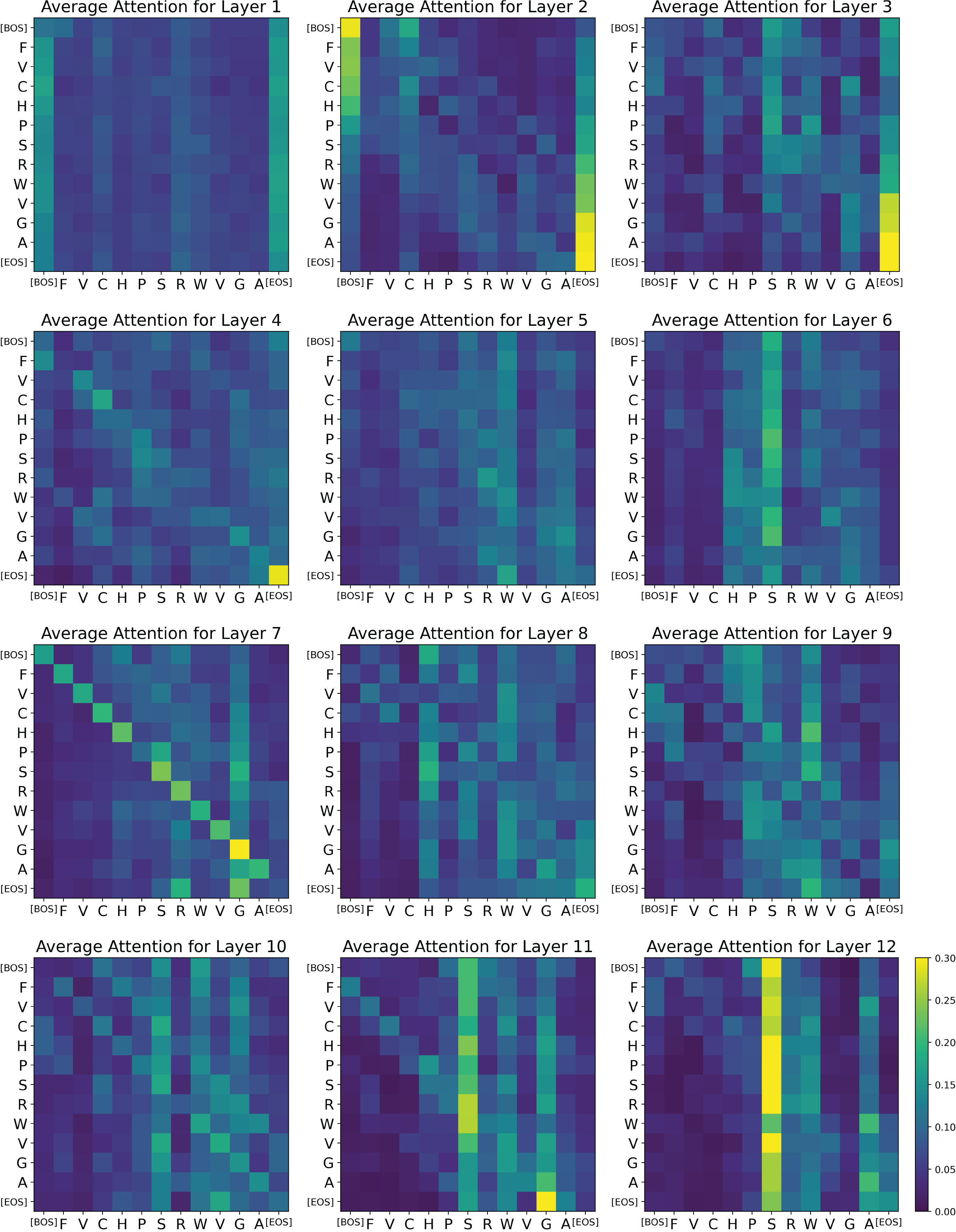}

     \caption{The average attention for all 12 layers of the fine-tuned LazBF-ESM for the LazBF substrate FVCHPSRWVGA.}
     \label{fig:figS9}
    \end{figure}


    \begin{table}
        \caption{The optimal hyperparameters for each downstream model type trained on each set of embeddings for the high-N condition.}
        \label{tbl:example}
        \begin{tabular}{|l|l|l|}
        \hline
        Embedding & Downstream Model & Hyperparameters \\
        \hline
        LazBF Vanilla-ESM  & SVC & C$=$5, kernel='rbf' \\
        LazBF Vanilla-ESM  & MLP & hidden\_layer\_sizes$=$1000, activation$=$'relu' \\
        LazBF Vanilla-ESM  & LR & C$=0.1$, penalty$=$'l2' \\
        LazBF Vanilla-ESM  & RF & n\_estimators$=$500, criterion$=$'log\_loss' \\
        LazBF Vanilla-ESM  & AB & learning\_rate$=$1, n\_estimators$=$500 \\
        LazBF Vanilla-ESM  & KNN & n\_neighbors$=$25, weights$=$'distance' \\
    
        LazDEF Vanilla-ESM  & SVC & C$=$5, kernel='rbf' \\
        LazDEF Vanilla-ESM  & MLP & hidden\_layer\_sizes$=$1000, activation$=$'relu' \\
        LazDEF Vanilla-ESM  & LR & C$=$0.1, penalty$=$'l2' \\
        LazDEF Vanilla-ESM  & RF & n\_estimators$=$500, criterion$=$'gini' \\
        LazDEF Vanilla-ESM  & AB & learning\_rate$=$1, n\_estimators$=$500 \\
        LazDEF Vanilla-ESM  & KNN & n\_neighbors$=$50, weights$=$'uniform' \\

    
    
        LazBF LazBF-ESM  & SVC & C$=$10, kernel='rbf' \\
        LazBF LazBF-ESM  & MLP & hidden\_layer\_sizes$=$1000, activation$=$'relu' \\
        LazBF LazBF-ESM  & LR & C$=$0.1, penalty$=$None \\
        LazBF LazBF-ESM  & RF & n\_estimators$=$100, criterion$=$'gini' \\
        LazBF LazBF-ESM  & AB & learning\_rate$=$0.1, n\_estimators$=$500 \\
        LazBF LazBF-ESM  & KNN & n\_neighbors$=$25, weights$=$'distance' \\
    
        LazDEF LazBF-ESM & SVC & C$=$5, kernel='rbf' \\
        LazDEF LazBF-ESM & MLP & hidden\_layer\_sizes$=$1000, activation$=$'relu' \\
        LazDEF LazBF-ESM & LR & C$=$0.1, penalty$=$'l2' \\
        LazDEF LazBF-ESM & RF & n\_estimators$=$500, criterion$=$'log\_loss' \\
        LazDEF LazBF-ESM & AB & learning\_rate$=$1, n\_estimators$=$500 \\
        LazDEF LazBF-ESM & KNN & n\_neighbors$=$50, weights$=$'distance' \\
    
        LazBF LazDEF-ESM & SVC & C$=$5, kernel='rbf' \\
        LazBF LazDEF-ESM & MLP & hidden\_layer\_sizes$=$1000, activation$=$'relu' \\
        LazBF LazDEF-ESM & LR & C$=$0.1, penalty$=$None \\
        LazBF LazDEF-ESM & RF & n\_estimators$=$500, criterion$=$'log\_loss' \\
        LazBF LazDEF-ESM & AB & learning\_rate$=$1, n\_estimators$=$500 \\
        LazBF LazDEF-ESM & KNN & n\_neighbors$=$50, weights$=$'uniform' \\
    
        LazDEF LazDEF-ESM & SVC & C$=$1, kernel='rbf' \\
        LazDEF LazDEF-ESM & MLP & hidden\_layer\_sizes$=$1000, activation$=$'relu' \\
        LazDEF LazDEF-ESM & LR & C$=$10, penalty$=$None \\
        LazDEF LazDEF-ESM & RF & n\_estimators$=$500, criterion$=$'gini' \\
        LazDEF LazDEF-ESM & AB & learning\_rate$=$1, n\_estimators$=$500 \\
        LazDEF LazDEF-ESM & KNN & n\_neighbors$=$50, weights$=$'uniform' \\
        \hline
    \end{tabular}
    \end{table}
    \begin{table}
        \caption{The optimal hyperparameters for each downstream model type trained on each set of embeddings for the medium-N condition.}
        \label{tbl:example}
        \begin{tabular}{|l|l|l|}
        \hline
        Embedding & Downstream Model & Hyperparameters \\
        \hline
        LazBF Vanilla-ESM  & SVC & C$=$10, kernel='linear' \\
        LazBF Vanilla-ESM  & MLP & hidden\_layer\_sizes$=$500, activation$=$'relu' \\
        LazBF Vanilla-ESM  & LR & C$=$10, penalty$=$None \\
        LazBF Vanilla-ESM  & RF & n\_estimators$=$200, criterion$=$'gini' \\
        LazBF Vanilla-ESM  & AB & learning\_rate$=$1, n\_estimators$=$200 \\
        LazBF Vanilla-ESM  & KNN & n\_neighbors$=$50, weights$=$'distance' \\
    
        LazDEF Vanilla-ESM  & SVC & C$=$1, kernel='rbf' \\
        LazDEF Vanilla-ESM  & MLP & hidden\_layer\_sizes$=$100, activation$=$'relu' \\
        LazDEF Vanilla-ESM  & LR & C$=$0.1, penalty$=$None \\
        LazDEF Vanilla-ESM  & RF & n\_estimators$=$200, criterion$=$'entropy' \\
        LazDEF Vanilla-ESM  & AB & learning\_rate$=$0.1, n\_estimators$=$500 \\
        LazDEF Vanilla-ESM  & KNN & n\_neighbors$=$50, weights$=$'uniform' \\

    
    
        LazBF LazBF-ESM  & SVC & C$=$0.1, kernel='linear' \\
        LazBF LazBF-ESM  & MLP & hidden\_layer\_sizes$=$50, activation$=$'tanh' \\
        LazBF LazBF-ESM  & LR & C$=$0.1, penalty$=$'l2' \\
        LazBF LazBF-ESM  & RF & n\_estimators$=$200, criterion$=$'entropy' \\
        LazBF LazBF-ESM  & AB & learning\_rate$=$0.1, n\_estimators$=$200 \\
        LazBF LazBF-ESM  & KNN & n\_neighbors$=$25, weights$=$'uniform' \\
    
        LazDEF LazBF-ESM & SVC & C$=$1, kernel='linear' \\
        LazDEF LazBF-ESM & MLP & hidden\_layer\_sizes$=$100, activation$=$'tanh' \\
        LazDEF LazBF-ESM & LR & C$=$0.1, penalty$=$'l2' \\
        LazDEF LazBF-ESM & RF & n\_estimators$=$500, criterion$=$'gini' \\
        LazDEF LazBF-ESM & AB & learning\_rate$=$1, n\_estimators$=$200 \\
        LazDEF LazBF-ESM & KNN & n\_neighbors$=$25, weights$=$'uniform' \\
    
        LazBF LazDEF-ESM & SVC & C$=$0.1, kernel='linear' \\
        LazBF LazDEF-ESM & MLP & hidden\_layer\_sizes$=$500, activation$=$'relu' \\
        LazBF LazDEF-ESM & LR & C$=$0.1, penalty$=$'l2' \\
        LazBF LazDEF-ESM & RF & n\_estimators$=$500, criterion$=$'log\_loss' \\
        LazBF LazDEF-ESM & AB & learning\_rate$=$1, n\_estimators$=$200 \\
        LazBF LazDEF-ESM & KNN & n\_neighbors$=$50, weights$=$'uniform' \\
    
        LazDEF LazDEF-ESM & SVC & C$=$1, kernel='rbf' \\
        LazDEF LazDEF-ESM & MLP & hidden\_layer\_sizes$=$100, activation$=$'relu' \\
        LazDEF LazDEF-ESM & LR & C$=$5, penalty$=$None \\
        LazDEF LazDEF-ESM & RF & n\_estimators$=$50, criterion$=$'entropy' \\
        LazDEF LazDEF-ESM & AB & learning\_rate$=$0.1, n\_estimators$=$500 \\
        LazDEF LazDEF-ESM & KNN & n\_neighbors$=$25, weights$=$'uniform' \\
        \hline
    \end{tabular}
    \end{table}
    
    \begin{table}
        \caption{The optimal hyperparameters for each downstream model type trained on each set of embeddings for the low-N condition.}
        \label{tbl:example}
        \begin{tabular}{|l|l|l|}
        \hline
        Embedding & Downstream Model & Hyperparameters \\
        \hline
        LazBF Vanilla-ESM  & SVC & C$=$5, kernel='rbf' \\
        LazBF Vanilla-ESM  & MLP & hidden\_layer\_sizes$=$100, activation$=$'relu' \\
        LazBF Vanilla-ESM  & LR & C$=$0.1, penalty$=$None \\
        LazBF Vanilla-ESM  & RF & n\_estimators$=$100, criterion$=$'entropy' \\
        LazBF Vanilla-ESM  & AB & learning\_rate$=$1, n\_estimators$=$500 \\
        LazBF Vanilla-ESM  & KNN & n\_neighbors$=$5, weights$=$'uniform' \\
    
        LazDEF Vanilla-ESM  & SVC & C$=$0.1, kernel='linear' \\
        LazDEF Vanilla-ESM  & MLP & hidden\_layer\_sizes$=$1000, activation$=$'relu' \\
        LazDEF Vanilla-ESM  & LR & C$=$0.1, penalty$=$'l2' \\
        LazDEF Vanilla-ESM  & RF & n\_estimators$=$100, criterion$=$'entropy' \\
        LazDEF Vanilla-ESM  & AB & learning\_rate$=$1, n\_estimators$=$200 \\
        LazDEF Vanilla-ESM  & KNN & n\_neighbors$=$10, weights$=$'uniform' \\

    
    
        LazBF LazBF-ESM  & SVC & C$=$0.1, kernel='rbf' \\
        LazBF LazBF-ESM  & MLP & hidden\_layer\_sizes$=$500, activation$=$'relu' \\
        LazBF LazBF-ESM  & LR & C$=$0.1, penalty$=$'l2' \\
        LazBF LazBF-ESM  & RF & n\_estimators$=$200, criterion$=$'entropy' \\
        LazBF LazBF-ESM  & AB & learning\_rate$=$5, n\_estimators$=$200 \\
        LazBF LazBF-ESM  & KNN & n\_neighbors$=$5, weights$=$'uniform' \\
    
        LazDEF LazBF-ESM & SVC & C$=$5, kernel='rbf' \\
        LazDEF LazBF-ESM & MLP & hidden\_layer\_sizes$=$750, activation$=$'relu' \\
        LazDEF LazBF-ESM & LR & C$=$0.1, penalty$=$'l2' \\
        LazDEF LazBF-ESM & RF & n\_estimators$=$200, criterion$=$'gini' \\
        LazDEF LazBF-ESM & AB & learning\_rate$=$1, n\_estimators$=$500 \\
        LazDEF LazBF-ESM & KNN & n\_neighbors$=$50, weights$=$'distance' \\
    
        LazBF LazDEF-ESM & SVC & C$=$0.1, kernel='linear' \\
        LazBF LazDEF-ESM & MLP & hidden\_layer\_sizes$=$750, activation$=$'relu' \\
        LazBF LazDEF-ESM & LR & C$=$0.1, penalty$=$'l2' \\
        LazBF LazDEF-ESM & RF & n\_estimators$=$50, criterion$=$'log\_loss' \\
        LazBF LazDEF-ESM & AB & learning\_rate$=$1, n\_estimators$=$500 \\
        LazBF LazDEF-ESM & KNN & n\_neighbors$=$10, weights$=$'uniform' \\
    
        LazDEF LazDEF-ESM & SVC & C$=$1, kernel='rbf' \\
        LazDEF LazDEF-ESM & MLP & hidden\_layer\_sizes$=$500, activation$=$'tanh' \\
        LazDEF LazDEF-ESM & LR & C$=$0.1, penalty$=$'l2' \\
        LazDEF LazDEF-ESM & RF & n\_estimators$=$200, criterion$=$'log\_loss' \\
        LazDEF LazDEF-ESM & AB & learning\_rate$=$0.1, n\_estimators$=$50 \\
        LazDEF LazDEF-ESM & KNN & n\_neighbors$=$50, weights$=$'uniform' \\
        \hline
    \end{tabular}
    \end{table}

    \begin{table}
        \caption{The optimal hyperparameters for each downstream model type trained on Peptide-ESM embeddings for the low-N, med-N, and high-N conditions.}
        \label{tbl:example}
        \begin{tabular}{|l|l|l|}
        \hline
        Embedding & Downstream Model & Hyperparameters \\
        \hline

        LazBF Peptide-ESM low-N  & SVC & C$=$0.1, kernel='linear' \\ 
        LazBF Peptide-ESM low-N  & MLP & hidden\_layer\_sizes$=$750, activation$=$'relu' \\
        LazBF Peptide-ESM low-N  & LR & C$=$1, penalty$=$None \\
        LazBF Peptide-ESM low-N  & RF & n\_estimators$=$50, criterion$=$'entropy' \\
        LazBF Peptide-ESM low-N  & AB & learning\_rate$=$1, n\_estimators$=$50 \\
        LazBF Peptide-ESM low-N  & KNN & n\_neighbors$=$10, weights$=$'uniform' \\
    
        LazDEF Peptide-ESM low-N  & SVC & C$=$1, kernel='linear' \\
        LazDEF Peptide-ESM low-N  & MLP & hidden\_layer\_sizes$=$500, activation$=$'relu' \\
        LazDEF Peptide-ESM low-N  & LR & C$=$1, penalty$=$'None' \\
        LazDEF Peptide-ESM low-N  & RF & n\_estimators$=$200, criterion$=$'gini' \\
        LazDEF Peptide-ESM low-N  & AB & learning\_rate$=$1, n\_estimators$=$200 \\
        LazDEF Peptide-ESM low-N  & KNN & n\_neighbors$=$25, weights$=$'uniform' \\

        LazBF Peptide-ESM med-N  & SVC & C$=$1, kernel='linear' \\ 
        LazBF Peptide-ESM med-N  & MLP & hidden\_layer\_sizes$=$500, activation$=$'tanh' \\
        LazBF Peptide-ESM med-N  & LR & C$=$1, penalty$=$None \\
        LazBF Peptide-ESM med-N  & RF & n\_estimators$=$500, criterion$=$'entropy' \\
        LazBF Peptide-ESM med-N  & AB & learning\_rate$=$0.1, n\_estimators$=$200 \\
        LazBF Peptide-ESM med-N  & KNN & n\_neighbors$=$10, weights$=$'uniform' \\
    
        LazDEF Peptide-ESM med-N  & SVC & C$=$0.1, kernel='linear' \\
        LazDEF Peptide-ESM med-N  & MLP & hidden\_layer\_sizes$=$1000, activation$=$'relu' \\
        LazDEF Peptide-ESM med-N  & LR & C$=$0.1, penalty$=$'None' \\
        LazDEF Peptide-ESM med-N  & RF & n\_estimators$=$200, criterion$=$'entropy' \\
        LazDEF Peptide-ESM med-N  & AB & learning\_rate$=$1, n\_estimators$=$500 \\
        LazDEF Peptide-ESM med-N  & KNN & n\_neighbors$=$25, weights$=$'distance' \\

        LazBF Peptide-ESM high-N  & SVC & C$=$0.1, kernel='linear' \\ 
        LazBF Peptide-ESM high-N  & MLP & hidden\_layer\_sizes$=$750, activation$=$'tanh' \\ 
        LazBF Peptide-ESM high-N  & LR & C$=$5, penalty$=$None \\
        LazBF Peptide-ESM high-N  & RF & n\_estimators$=$500, criterion$=$'entropy' \\
        LazBF Peptide-ESM high-N  & AB & learning\_rate$=$1, n\_estimators$=$500 \\
        LazBF Peptide-ESM high-N  & KNN & n\_neighbors$=$50, weights$=$'uniform' \\
    
        LazDEF Peptide-ESM high-N  & SVC & C$=$5, kernel='linear' \\ 
        LazDEF Peptide-ESM high-N  & MLP & hidden\_layer\_sizes$=$50, activation$=$'relu' \\ 
        LazDEF Peptide-ESM high-N  & LR & C$=$1, penalty$=$'None' \\
        LazDEF Peptide-ESM high-N  & RF & n\_estimators$=$500, criterion$=$'log loss' \\
        LazDEF Peptide-ESM high-N  & AB & learning\_rate$=$1, n\_estimators$=$500 \\
        LazDEF Peptide-ESM high-N  & KNN & n\_neighbors$=$50, weights$=$'uniform' \\
    
        \hline
    \end{tabular}
    \end{table}